\documentclass[12pt,a4paper]{article}

\usepackage[font=small]{caption}
\usepackage[text={450pt,650pt},headheight={14.5pt},centering]{geometry}

\usepackage{lmodern}
\usepackage{cite,hyperref}
\usepackage{graphicx}
\usepackage{booktabs}
\usepackage{subfig}
\usepackage{tikz} %tikz figures
\usetikzlibrary{calc}
\usetikzlibrary{arrows}

\usepackage{amsmath,amssymb}
\usepackage{mathrsfs}
\usepackage{mathtools}
\usepackage{bbm}
\usepackage{slashed}
\usepackage{braket}

\usepackage{dsfont}

\usepackage{accents}

\DeclareMathAlphabet{\mathsfit}{\encodingdefault}{\sfdefault}{m}{sl}

\usepackage{xspace}

\numberwithin{equation}{section}

\makeatletter
 \let\old@startsection=\@startsection
 \let\oldl@section=\l@section
 \renewcommand{\@startsection}[6]{\old@startsection{#1}{#2}{#3}{#4}{#5}{#6\mathversion{bold}}}
 \renewcommand{\l@section}[2]{\oldl@section{\mathversion{bold}#1}{#2}}
\makeatother

\renewcommand{\leq}{\leqslant}

\def\XXint#1#2#3{{\setbox0=\hbox{$#1{#2#3}{\int}$}
    \vcenter{\hbox{$#2#3$}}\kern-.5\wd0}}

\newcommand{\ce}{\text{c.e.}}

\newcommand{\AdS}{\textup{AdS}}
\newcommand{\CFT}{\textup{CFT}}
\newcommand{\Sphere}{\textup{S}}
\newcommand{\Torus}{\textup{T}}
\newcommand{\CP}{\Complex \textup{P}}

\newcommand{\Smat}{\mathcal{S}}

\newcommand{\alg}[1]{\mathrm{#1}}
\newcommand{\grp}[1]{\mathrm{#1}}

\newcommand{\gen}[1]{\mathbf{#1}}

\newcommand{\genQ}{\gen{Q}}

\newcommand{\genH}{\gen{H}}

\newcommand{\genM}{\gen{M}}

\newcommand{\Integers}{\mathbbm{Z}}

\newcommand{\Complex}{\mathbbm{C}}

\newcommand{\so}{\alg{so}}

\newcommand{\su}{\alg{su}}
\newcommand{\psu}{\alg{psu}}

\newcommand{\algD}[1]{\alg{d}(2,1;#1)}
\newcommand{\grpD}[1]{\grp{D}(2,1;#1)}

\newcommand{\algSL}{\alg{sl}}
\newcommand{\grpSL}{\grp{SL}}

\newcommand{\algSU}{\alg{su}}

\newcommand{\algU}{\alg{u}}

\newcommand{\algPSU}{\alg{psu}}
\newcommand{\grpPSU}{\grp{PSU}}

\newcommand{\grpOSp}{\grp{OSp}}

\newcommand{\wh}{\hat{\omega}}
\newcommand{\wc}{\check{\omega}}

\newcommand{\pri}[1]{\accentset{\prime}{#1}}

\newcommand{\superN}{\mathcal{N}}

\newcommand{\ie}{\textit{i.e.}\xspace}
\newcommand{\eg}{\textit{e.g.}\xspace}

\newcommand{\acomm}[2]{\{#1,#2\}}
\newcommand{\acommPB}[2]{\{#1,#2\}_{\mbox{\tiny PB}}}

\newcommand{\sL}{\mbox{\tiny L}}
\newcommand{\sR}{\mbox{\tiny R}}

\newcommand{\smallL}{\sL}
\newcommand{\smallR}{\sR}

\newcommand{\atoone}{\xrightarrow{\alpha \to 1}}
\newcommand{\QL}{\genQ_{\sL}}
\newcommand{\QR}{\genQ_{\sR}}
\newcommand{\QbL}{\overline{\genQ}_{\sL}}
\newcommand{\QbR}{\overline{\genQ}_{\sR}}

\newcommand{\hslashslash}{%
  \raisebox{.9ex}{%
    \scalebox{.7}{%
      \rotatebox[origin=c]{17}{$-$}%
    }%
  }%
}
\renewcommand{\k}{%
  {%
   \vphantom{k}%
   \ooalign{\kern-.05em\smash{\hslashslash}\hidewidth\cr$k$\cr}%
   \kern-.025em
  }%
}

\newcommand{\h}{h}

\newcommand{\algA}{\mathcal{A}}

\begin{document}

\thispagestyle{empty}

\begin{flushright}\footnotesize\ttfamily
ITP-UU-15/08
\\
HU-EP-15/26
\\
HU-Mathematik-P-2015-06
\\
Imperial-TP-OOS-2015-01
\\
\end{flushright}
\vspace{5em}

\begin{center}
\textbf{\Large\mathversion{bold} The $\AdS_3\times \Sphere^3\times \Sphere^3\times\Sphere^1 $ worldsheet S matrix}

\vspace{2em}

\textrm{\large Riccardo Borsato${}^1$, Olof Ohlsson Sax${}^2$, Alessandro Sfondrini${}^{3}$\\
and Bogdan Stefa\'nski jr.${}^4$ } 

\vspace{2em}

\begingroup\itshape
1. Institute for Theoretical Physics and Spinoza Institute, Utrecht University,\\ Leuvenlaan 4, 3584 CE Utrecht, The Netherlands\\[0.2cm]

2. The Blackett Laboratory, Imperial College,\\  SW7 2AZ, London,
U.K.\\[0.2cm]

3. Institut f\"ur Mathematik und Institut f\"ur Physik, Humboldt-Universit\"at zu Berlin\\
IRIS Geb\"aude, Zum Grossen Windkanal 6, 12489 Berlin, Germany\\[0.2cm]

4. Centre for Mathematical Science, City University London,\\ Northampton Square, EC1V 0HB, London, U.K.\par\endgroup

\vspace{1em}

\texttt{R.Borsato@uu.nl, o.olsson-sax@imperial.ac.uk, Alessandro.Sfondrini@physik.hu-berlin.de, Bogdan.Stefanski.1@city.ac.uk}

%%%%%%%%

\end{center}

\vspace{6em}

\begin{abstract}\noindent
We investigate type IIB strings on $\AdS_3\times \Sphere^3\times \Sphere^3\times\Sphere^1 $ with mixed Ramond-Ramond (R-R) and Neveu-Schwarz-Neveu-Schwarz (NS-NS) flux. By suitably gauge-fixing the closed string Green-Schwarz (GS) action of this theory, we derive the off-shell symmetry algebra and its representations. We use these to determine the non-perturbative worldsheet S-matrix of fundamental excitations in the theory. The analysis involves both massive and massless modes in complete generality. The S-matrix we find involves a number of phase factors, which in turn satisfy crossing equations that we also determine. We comment on the nature of the heaviest modes of the theory,  but leave their identification either as composites or bound-states to a future investigation.
\end{abstract}

%%%%%%%%%%%%%%%%%%%%%%%%%%%%%%%%%%%%%%%%%%%%%%%%%%%%%%%%%%%%%%%%%%%%%%%%%%%
\newpage

\tableofcontents

\section{Introduction}
\label{sec:introduction}

The holographic correspondence between gravity and quantum field theories~\cite{'tHooft:1993gx} can be quantitatively realised in string theory as a duality between superstrings on anti-De Sitter (AdS) space and conformal field theories (CFT)~\cite{Maldacena:1997re,Witten:1998qj,Gubser:1998bc}.
As this $\AdS/\CFT$ correspondence is a weak-strong duality, it is highly desirable to find exact approaches to study it. In the 't~Hooft, or planar, limit~\cite{'tHooft:1973jz} of certain classes of dual theories a very successful approach is \emph{integrability}---finding hidden symmetries that allow for the solution of the spectrum of protected \emph{and non-protected} states of both theories. 
The best understood $\AdS/\CFT$ dual pairs are given by type IIB strings on~$\AdS_5\times \Sphere^5$ and the dual $\mathcal{N}=4$ Supersymmetric Yang-Mills (SYM) theory, and its close relative type IIA string theory on $\AdS_4\times \mathbb{C}\text{P}^3$~\cite{Stefanski:2008ik,Arutyunov:2008if,Gomis:2008jt} and the dual ABJM Chern-Simons theory~\cite{Bagger:2007jr,Aharony:2008ug}, see references~\cite{Arutyunov:2009ga,Beisert:2010jr,Klose:2010ki} for reviews and a more complete list of references. Integrability seems to be quite a robust feature of such backgrounds, as it persists for their orbifolds, orientifolds as well as for certain deformations~\cite{Zoubos:2010kh,vanTongeren:2013gva}. It is natural to wonder if integrability underlies other instances of AdS/CFT, and in particular whether $\AdS_3/\CFT_2$ enjoys such hidden symmetries.

It turns out that superstrings on~$\AdS_3\times \mathcal{M}_7$ with the maximal amount of supersymmetry allowed for such backgrounds (16 real supercharges)~\cite{Pesando:1998wm, Rahmfeld:1998zn,Park:1998un, Metsaev:2000mv,Babichenko:2009dk} are indeed classically integrable~\cite{Babichenko:2009dk,Sundin:2012gc}. More precisely, the classical superstring non-linear sigma model on 
\begin{equation}
\AdS_3\times\Sphere^3\times\Torus^4\qquad\text{and}\qquad
\AdS_3\times\Sphere^3\times\Sphere^3\times\Sphere^1 
\end{equation}
supported by R-R background fluxes admits a Lax formulation. In fact, such $\AdS_3$ backgrounds supported by a {\em mixture} of R-R and NS-NS three-form fluxes are integrable~\cite{Cagnazzo:2012se}. These results indicate that integrability may underlie the $\AdS_3/\CFT_2$ correspondence, but are not enough to determine whether the spectrum of the quantum theory can be found by Bethe ansatz techniques. In this paper, we construct an S~matrix for the scattering of asymptotic excitations on the string worldsheet, that is compatible with the assumption of quantum integrability, in particular with factorised scattering. In this context the scattering of giant magnons in $\AdS_3$ was originally investigated in~\cite{David:2008yk,David:2010yg} and more recently in~\cite{Stepanchuk:2014kza}. In this paper we will construct the worldsheet S~matrix for the $\AdS_3\times\Sphere^3\times\Sphere^3\times\Sphere^1$ background by studying the off-shell symmetry algebra of the light-cone gauge-fixed string theory. In the case of $\AdS_5/\CFT_4$ correspondence constraining the S matrix by the off-shell symmetry algebra was first developed in the spin-chain setting in~\cite{Beisert:2005tm}. On the string theory side a corresponding derivation of the $\AdS_5\times\Sphere^5$ worldsheet S matrix was done in~\cite{Arutyunov:2006ak, Arutyunov:2006yd} and applied to $\AdS_3\times\Sphere^3\times\Torus^4$ in~\cite{Borsato:2014exa, Borsato:2014hja, Lloyd:2014bsa}.

This method circumvents the problems associated with the presence of \emph{massless} worldsheet excitations typically found in $\AdS_3\times \mathcal{M}_7$ backgrounds. Considerable progress had been made in the study of \emph{massive} modes
on  $\AdS_3\times\Sphere^3\times\Sphere^3\times\Sphere^1$~\cite{OhlssonSax:2011ms, Borsato:2012ud,Borsato:2012ss,Abbott:2012dd, Beccaria:2012kb,Beccaria:2012pm,Abbott:2013ixa}.%
\footnote{%
Massive modes on $\AdS_3\times\Sphere^3\times\Torus^4$ were understood
in a similar manner~\cite{Borsato:2013qpa,Borsato:2013hoa}.  See~\cite{Sfondrini:2014via} for a review and more extensive list of references.
}
In particular, the all-loop massive S~matrix~\cite{Borsato:2012ud} and Bethe ansatz~\cite{Borsato:2012ss} were found in the background supported by pure R-R flux up to the so-called dressing factors. It was harder to incorporate fully the massless modes into the integrable structure, though partial progress in this direction 
was made in~\cite{Sax:2012jv,Lloyd:2013wza,Abbott:2014rca}. To date, no proposal existed for scattering of massless modes in $\AdS_3\times\Sphere^3\times\Sphere^3\times\Sphere^1$, nor the inclusion of NS-NS flux.\footnote{For recent work on integrable $\AdS_3$ string solutions involving NS-NS flux see~\cite{Ahn:2014tua,Babichenko:2014yaa,
David:2014qta,Hernandez:2014eta,Hernandez:2015nba}.}

The methods employed in this paper naturally incorporate both massive and massless modes and allows for mixed R-R and NS-NS fluxes. The starting point is the GS action of type IIB string theory on $\AdS_3\times\Sphere^3\times\Sphere^3\times\Sphere^1$ with mixed flux. Strings in this background possess a \emph{large} $(4,4)$ super-conformal algebra~\cite{Sevrin:1988ew}, whose finite-dimensional sub-algebra is $\alg{d}(2,1;\alpha)^2$~\cite{Gauntlett:1998kc}. Upon gauge-fixing the GS action, only a sub-algebra
\begin{equation}
\alg{su}(1|1)^2\subset\alg{d}(2,1;\alpha)^2
\end{equation}
commutes with the Hamiltonian. When the level-matching condition is relaxed 
$\alg{su}(1|1)^2$ acquires two new central charges $\gen{C},\overline{\gen{C}}$. We denote this off-shell symmetry algebra by $\algA$.\footnote{The appearance of such central extensions when level-matching is relaxed is similar to what happens in $\AdS_5\times\Sphere^5$~\cite{Arutyunov:2006ak, Arutyunov:2006yd} and  $\AdS_3\times\Sphere^3\times \Torus^4$~\cite{Borsato:2014exa, Borsato:2014hja, Lloyd:2014bsa}.} The world sheet S-matrix  of the theory can be fixed, up to dressing phases, by requiring that it commute with $\algA$. In this paper we write down the world sheet S-matrix of this theory and show that it satisfies the Yang-Baxter equation. We also determine the crossing equations that the dressing phases have to satisfy. In this way, we find evidence for a family of integrable theories interpolating between the pure R-R-flux case familiar from $\AdS/\CFT$ in higher dimensions and the pure NS-NS case which is well-understood through worldsheet CFT techniques~\cite{Maldacena:2000hw}.

This paper is structured as follows. In section~\ref{sec:GS} we derive the algebra $\algA$
from a gauge-fixed GS action of strings on $\AdS_3\times\Sphere^3\times\Sphere^3\times\Sphere^1$.  In section~\ref{sec:quadraticrepr}  we study the representations of $\mathcal{A}$ at quadratic order in fields. We comment on the possible interpretation of the heavy modes as composite modes or bound states of the theory, and the consequences this would have. We leave the question of determining the exact nature of these modes to future investigations.  In section~\ref{sec:exact-rep} we write down the exact representations of $\algA$. In section~\ref{sec:int-s-matrix} we use these representations and the off-shell form of the algebra $\algA$ to fix the structure of the two-body worldsheet S-matrix up to a number of dressing phases. Using unitarity and crossing, we reduce the number of independent dressing phases and determine the crossing equations that these phases have to satisfy. Following our conclusions, we include a number of technical appendices. 

In much of section~\ref{sec:GS} we write down expressions that are leading order in fermionic fields and next-to-leading order in bosonic fields. We have used a Mathematica program to find these expressions and we include the program as part of our submission. The program contains expressions which are next-to-next-to-leading order in bosonic fields. These expressions are very lengthy and we have not transferred them to the present manuscript. The interested reader may find them by running the Mathematica program. We have nonetheless checked that the derivation of the centrally-extended algebra $\mathcal{A}$  remains valid at this order in the bosonic fields. The Mathematica package \texttt{grassmann.m} by M. Headrick and J. Michelson was very useful when performing the calculations presented in the first part of this paper.

\section{String theory on \texorpdfstring{$\AdS_3 \times \Sphere^3 \times \Sphere^3 \times \Sphere^1$}{AdS3 x S3 x S3 x S1} and the off-shell symmetry algebra}
\label{sec:GS}

In this section we write down the fully gauge-fixed Green-Schwarz action for type IIB string theory on $\AdS_3 \times \Sphere^3 \times \Sphere^3 \times \Sphere^1$ with mixed flux up to quadratic order in fermions. We determine the classical conserved supercharges of the theory and calculate the off-shell algebra $\mathcal{A}$ that they satisfy.

\subsection{The supergravity background}

We write the metric of $\AdS_3 \times \Sphere^3 \times \Sphere^3 \times \Sphere^1$ as
\begin{equation}
  ds^2 = ds_{\AdS_3}^2 + ds_{\Sphere^3_+}^2 + ds_{\Sphere^3_-}^2 + dw^2 ,
\end{equation}
where $w$ is the coordinate along the $\Sphere^1$. The radii of $\AdS_3$ and of the two three-spheres are related by~\cite{Gauntlett:1998kc}
\begin{equation}
  \frac{1}{R_{\AdS_3}^2} = \frac{1}{R_{\Sphere^3_+}^2} + \frac{1}{R_{\Sphere^3_-}^2} .
\end{equation}
We normalise the $\AdS_3$ radius to one and solve the above relation by setting
\begin{equation}
  \frac{1}{R_{\Sphere^3_+}^2} = \alpha \equiv \cos^2\!\varphi , \qquad
  \frac{1}{R_{\Sphere^3_-}^2} = 1-\alpha \equiv \sin^2\!\varphi .
\end{equation}
The metrics on $\AdS_3$ and the spheres are then given by\footnote{%
  The coordinates of the three-spheres have been rescaled by the radius of respective sphere, so that for example the angle $\phi_8$ takes values $0 \le \phi_8 < 2\pi R_{\Sphere_3^-} = 2\pi / \sin\varphi$ and $(x_6,x_7)$ take values on a disc of radius $2R_{\Sphere_3^-} = 2/\sin\varphi$. This makes the expressions for the metric and $B$ field more complicated, but gives canonically normalised kinetic terms in the bosonic action and makes the limit $\varphi \to 0$, or $R_{\Sphere_3^-} \to \infty$, more straightforward.%
}
\begin{equation}\label{eq:metrics}
  \begin{aligned}
    ds^2_{\AdS^3} &= -\Bigl(\frac{1 + \frac{z_1^2 + z_2^2}{4}}{1 - \frac{z_1^2 + z_2^2}{4}}\Bigr)^2 dt^2 + \Bigl(\frac{1}{1 - \frac{z_1^2 + z_2^2}{4}}\Bigr)^2 ( dz_1^2 + dz_2^2 ), \\
    ds^2_{\Sphere^3_+} &= \Bigl(\frac{1 - \cos^2\!\varphi\, \frac{y_3^2 + y_4^2}{4}}{1 + \cos^2\!\varphi\, \frac{y_3^2 + y_4^2}{4}}\Bigr)^2 d\phi_5^2 + \Bigl(\frac{1}{1 + \cos^2\!\varphi\, \frac{y_3^2 + y_4^2}{4}}\Bigr)^2 ( dy_3^2 + dy_4^2 ), \\
    ds^2_{\Sphere^3_-} &= \Bigl(\frac{1 - \sin^2\!\varphi\, \frac{x_6^2 + x_7^2}{4}}{1 + \sin^2\!\varphi\, \frac{x_6^2 + x_7^2}{4}}\Bigr)^2 d\phi_8^2 + \Bigl(\frac{1}{1 + \sin^2\!\varphi\, \frac{x_6^2 + x_7^2}{4}}\Bigr)^2 ( dx_6^2 + dx_7^2 ).
  \end{aligned}
\end{equation}
The bosonic background further contains a $B$-field
\begin{equation}
  \begin{aligned}
    B =&
    \frac{q}{\bigl(1-\frac{z_1^2 + z_2^2}{4}\bigr)^2} \bigl( z_1 dz_2 - z_2 dz_1 \bigr) \wedge dt \\
    &+
    \frac{q \cos\!\varphi}{\bigl(1+\cos\!\varphi\frac{y_3^2 + y_4^2}{4}\bigr)^2} \bigl( y_3 dy_4 - y_4 dy_3 \bigr) \wedge d\phi_5 \\
    &+
    \frac{q \sin\!\varphi}{\bigl(1+\sin\!\varphi\frac{x_6^2 + x_7^2}{4}\bigr)^2} \bigl( x_6 dx_7 - x_6 dx_7 \bigr) \wedge d\phi_8 ,
  \end{aligned}
\end{equation}
where the parameter $q$ is related to the quantised coefficient $k$ of the Wess-Zumino (WZ) term by
\begin{equation}
k = q \sqrt{\lambda}.
\end{equation}
The corresponding NS-NS three form is given by
\begin{equation}
  H = dB = 2q\Bigl( \operatorname{Vol}(\AdS_3) + \frac{1}{\cos^2\varphi} \operatorname{Vol}(\Sphere^3_+) + \frac{1}{\sin^2\varphi} \operatorname{Vol}(\Sphere^3_-) \Bigr) ,
\end{equation}
where the volume forms are all defined for unit radius. For $q=1$ this precisely corresponds to an $\algSL(2)_k \times \algSU(2)_{k'} \times \algSU(2)_{k''}$ Wess-Zumino-Witten (WZW) model where the three levels satisfy~\cite{Elitzur:1998mm}.
\begin{equation}
  \frac{1}{k} = \frac{1}{k'} + \frac{1}{k''} .
\end{equation}
In addition to the NS-NS three form, the background contains a R-R three form
\begin{equation}
  F = 2\tilde{q}\Bigl( \operatorname{Vol}(\AdS_3) + \frac{1}{\cos^2\varphi} \operatorname{Vol}(\Sphere^3_+) + \frac{1}{\sin^2\varphi} \operatorname{Vol}(\Sphere^3_-) \Bigr) \,,
\end{equation}
where
\begin{equation}
\tilde{q}=\sqrt{1-q^2}\,.
\end{equation}
In appendix~\ref{sec:Killing-spinors} we write down the Killing spinors for this background.

\subsection{Bosonic action and gauge fixing}

The action for the bosonic sigma model is given by\footnote{%
  In writing down the action and supercurrents in this section we suppress the string tension $\sqrt{\lambda}/2\pi$. We will reinstate in the relevant places in the next section.%
}%
\begin{equation}
  S_B = -\frac{1}{2} \int d\sigma d\tau \bigl(
  \gamma^{\alpha\beta} G_{MN} \partial_{\alpha} X^M \partial_{\beta} X^N + \epsilon^{\alpha\beta} B_{MN} \partial_{\alpha} X^M \partial_{\beta} X^N
  \bigr) .
\end{equation}
Introducing the canonically conjugate momenta
\begin{equation}
  p_M = \frac{\delta S_B}{\delta\dot{X}^M} = -\gamma^{0\beta} G_{MN} \partial_\beta X^N - B_{MN} \pri{X}^N 
\end{equation}
the bosonic action can be written in the first order form
\begin{equation}
  S_B = \int d\sigma \bigl( p_M \dot{X}^M + \frac{\gamma^{01}}{\gamma^{00}} C_1 + \frac{1}{2\gamma^{00}} C_2 \bigr)
\end{equation}
with
\begin{equation}
  \begin{gathered}
    C_1 = p_M \pri{X}^M , \\
    C_2 = G^{MN} p_M p_N + G_{MN} \pri{X}^M \pri{X}^N + 2G^{MN} B_{NK} p_M \pri{X}^K + G^{MN} B_{MK} B_{NL} \pri{X}^K \pri{X}^L .
  \end{gathered}
\end{equation}
Above $\dot{}$ and $'$ denote derivatives with respect to $\tau$ and $\sigma$, respectively. We further introduce light-cone coordinates $x^{\pm}$ along the supersymmetric geodesic and a transverse angle $\psi$ by setting
\begin{equation}
  \begin{aligned}
    x^{\pm} &= \frac{1}{2} \bigl( \cos\varphi\,\phi_5 + \sin\varphi\,\phi_8 \pm t \bigr) , &
    \psi &= -\sin\varphi\,\phi_5 + \cos\varphi\,\phi_8 .
  \end{aligned}
\end{equation}
To fix uniform light-cone gauge we now set
\begin{equation}
  x^+ = \tau , \qquad
  p_- = 2 ,
\end{equation}
where $p_-$ is the canonical momentum conjugate to $x^-$. This completely fixes the dynamics of the light-cone directions $x^{\pm}$. The resulting gauge-fixed bosonic action can then be expanded in the eight remaining transverse fields.

The constraints $C_1=0$ and $C_2=0$ are equivalent to the Virasoro constraints 
\begin{equation}
  \begin{aligned}
    \gamma^{11} G_{MN} \dot{X}^M \pri{X}^N + \gamma^{01} G_{MN} \dot{X}^M \dot{X}^N &= 0 , \\
    \gamma^{00} G_{MN} \dot{X}^M \dot{X}^N - \gamma^{11} G_{MN} \pri{X}^M \pri{X}^N &= 0.
  \end{aligned}
\end{equation}
To cubic order in the transverse fields the worldsheet metric is then given by
\begin{equation}
  \begin{aligned}
    \gamma^{\tau\tau}
    &= -1 + \tfrac{1}{2} \bigl( z^2 - \cos^4\varphi \, y^2 - \sin^4\varphi \, x^2 \bigr)
    + \tfrac{1}{4} \sin(2\varphi) \, \dot{\psi} \bigl( \cos^2\varphi \, y^2 - \sin^2\varphi \, x^2 \bigr) ,
    \\
    \gamma^{\sigma\sigma}
    &= +1 + \tfrac{1}{2} \bigl( z^2 - \cos^4\varphi \, y^2 - \sin^4\varphi \, x^2 \bigr)
    + \tfrac{1}{4} \sin(2\varphi) \, \dot{\psi} \bigl( \cos^2\varphi \, y^2 - \sin^2\varphi \, x^2 \bigr) ,
    \\
    \gamma^{\tau\sigma}
    &= \hphantom{-1 + \tfrac{1}{2} \bigl( z^2 - \cos^4\varphi \, y^2 - \sin^4\varphi \, x^2 \bigr)}
    - \tfrac{1}{4} \sin (2\varphi) \, \pri{\psi} \bigl( \cos^2\varphi \, y^2 - \sin^2\varphi \, x^2 \bigr) .
  \end{aligned}
\end{equation}
The worldsheet derivatives of the light-cone coordinate $x^-$ can be found by imposing equations of motion and the gauge-fixing condition. To cubic order we find
\begin{equation}
  \begin{aligned}
    \pri{x}^-
    &=
    - \tfrac{1}{2} \bigl( \dot{z}_i \pri{z}_i + \dot{y}_i \pri{y}_i + \dot{x}_i \pri{x}_i + \dot{w} \pri{w} + \dot{\psi} \pri{\psi} \bigr)
    - \tfrac{1}{4} \sin(2\varphi) \, \pri{\psi} \bigl( \cos^2\varphi \, y^2 - \sin^2\varphi \, x^2 \bigr) ,
    \\
    \dot{x}^-
    &=
    -\tfrac{1}{4} \bigl( 
    \dot{z}^2 + \dot{y}^2 + \dot{x}^2 + \dot{w}^2 + \dot{\psi}^2
    + \pri{z}^2 + \pri{y}^2 + \pri{x}^2 + \pri{w}^2 + \pri{\psi}^2
    \\ &\qquad
    - z^2 - \cos^4\!\varphi \, y^2 - \sin^4\!\varphi \, x^2
    \bigr)
    + \tfrac{1}{4} \sin(2\varphi) \, \dot{\psi} \bigl( \cos^2\varphi \, y^2 - \sin^2\varphi \, x^2 \bigr) .
  \end{aligned}
\end{equation}
Because of the gauge fixing condition $p_- = 2$, the total light-cone momentum $P_-$ is given by
\begin{equation}
  P_- = \int_{-r}^{+r} p_- = 4r ,
\end{equation}
where we have introduced the integration limits $\pm r$ to keep track of the extent of the worldsheet. We will work in the large $P_-$ limit, where $r \to \infty$ and the worldsheet decompactifies and we are effectively on a plane rather than a cylinder. However, we still impose periodic boundary conditions on the fields. The field $x^+$ is independent of $\sigma$ and hence periodic. Imposing periodicity of $x^-$ we find the condition
\begin{equation}
  \Delta x^- = x^-(+\infty) - x^-(-\infty) = \int_{-\infty}^{+\infty} d\sigma \pri{x}^- = 0.
\end{equation}
From the constraint $C_1 = 0$ and the gauge fixing conditions we find
\begin{equation}
  2 \pri{x}^- = - \bigl( p_{z^i} \pri{z}^i + p_{y^i} \pri{y}^i + p_{x^i} \pri{x}^i + p_{w^i} \pri{w}^i + p_{\psi^i} \pri{\psi}^i \bigr) .
\end{equation}
The right-hand-side of the above expression is exactly the \emph{world sheet momentum density}. Hence,
\begin{equation}
  \Delta x^- = \frac{1}{2} p_{\text{w.s.}} .
\end{equation}
Above we have assumed that there is no winding along the direction $\phi$. In the general case, periodicity of $x^-$ gives the condition
\begin{equation}
  p_{\text{w.s.}} = 2\pi m ,
\end{equation}
where $m$ is the winding number.

\subsection{Green-Schwarz action and suitable fermionic coordinates}
\label{sec:GS-action-ferm-coords}

Having found the gauge fixing conditions from the bosonic action we will now write down the fermionic part of the GS action. The procedure here is very similar to the case of mixed flux $\AdS_3 \times \Sphere^3 \times \Torus^4$~\cite{Lloyd:2014bsa}.

The GS action is given by
\begin{equation}
\mathcal{L}=\mathcal{L}_{\text{B}}+\mathcal{L}_{\text{kin}}
+\mathcal{L}_{\text{WZ}},
\end{equation}
where $\mathcal{L}_{\text{B}}$ is the bosonic part of the action discussed in the previous sub-section and, up to quadratic order in fermions~\cite{Grisaru:1985fv,Cvetic:1999zs,Wulff:2013kga}
\begin{align}
  \mathcal{L}_{\text{kin}} &= 
  -i\gamma^{\alpha\beta} \bar{\tilde{\theta}}_I \slashed{E}_\alpha \bigl( \delta^{IJ} D_\beta + \frac{1}{48} \sigma_3^{IJ} \slashed{F} \slashed{E}_\beta + \frac{1}{8} \sigma_1^{IJ} \slashed{H}_\beta \bigr) \tilde{\theta}_J  ,\\
  \mathcal{L}_{\text{WZ}} &= 
  +i\epsilon^{\alpha\beta} \bar{\tilde{\theta}}_I \sigma_1^{IJ} \slashed{E}_\alpha \bigl( \delta^{JK} D_\beta + \frac{1}{48} \sigma_3^{JK} \slashed{F} \slashed{E}_\beta + \frac{1}{8} \sigma_1^{JK} \slashed{H}_\beta \bigr) \tilde{\theta}_K .
\end{align}
Above, the fermions have beed ``rotated'' along the $I-J$ index compared to the expressions given in~\cite{Cvetic:1999zs}
\begin{equation}
\label{eq:fermion-IJ-rot}
  \tilde{\theta}_1 = \sqrt{\frac{1+\tilde{q}}{2}} \, \theta_1 - \sqrt{\frac{1-\tilde{q}}{2}} \, \theta_2  , \qquad
  \tilde{\theta}_2 = \sqrt{\frac{1+\tilde{q}}{2}} \, \theta_2 + \sqrt{\frac{1-\tilde{q}}{2}} \, \theta_1 .
\end{equation}
This ensures that the kinetic term in the Lagrangian is diagonal in terms of the $\theta_I$.

To understand the action of the supersymmetries on the fields it is useful to perform a field redefinition so that the fermions in the action are closely related to the Killing spinors of the background. We introduce the rotated fermions
\begin{equation}\label{eq:rotated-fermions-theta}
  \begin{aligned}
    \tilde{\theta}_1 &= \sqrt{\frac{1+\tilde{q}}{2}} M_0 \theta_1 - \sqrt{\frac{1-\tilde{q}}{2}} M_0^{-1} \theta_2 , \\
    \tilde{\theta}_2 &= \sqrt{\frac{1-\tilde{q}}{2}} M_0 \theta_1 + \sqrt{\frac{1+\tilde{q}}{2}} M_0^{-1} \theta_2 ,
  \end{aligned}
\end{equation}
where the matrix $M_0$ is given in equation~\eqref{eq:M0-Mt-definition}. To make the connection with Killing spinors manifest, we use the projectors
\begin{equation}
  \Pi_{\pm} = \frac{1}{2} ( 1 \pm \cos\varphi \, \Gamma^{012345} \pm \sin\varphi \, \Gamma^{012678} ) ,
\end{equation}
to further define
\begin{equation}
  \theta_1 = M_t \bigl( \Pi_{+} \vartheta_1^+ + \Pi_{-} \vartheta_1^- \bigr) , \qquad
  \theta_2 = M_t^{-1} \bigl( \Pi_{+} \vartheta_2^+ + \Pi_{-} \vartheta_2^- \bigr) ,
\end{equation}
where the matrix $M_t$ is given in equation~\eqref{eq:M0-Mt-definition}.
The action of the sixteen supersymmetries of $\algD{\alpha}^2$ then correspond to shifts in the fermions $\vartheta_I^-$. 

The GS action has a large gauge invariance. We fix this by a suitable choice of kappa and light-cone gauge. In uniform light-cone gauge, the directions $x^{\pm}$ play a special role. Under shifts of these light-cone coordinates the fermions $\vartheta_I^{\pm}$ change by a phase. In the gauge-fixed action it is therefore more convenient to use the fields $\theta_I$, which are neutral under such shifts.\footnote{%
  The fermions $\theta_I$ are also invariant under shifts of $\psi$. This is not essential for our calculation, but still convenient. Since the field $\psi$ is massless the action is invariant under shifts of $\psi$. However, if the fermions transform under such shifts there will be terms in the Lagrangian that depend on the field $\psi$ itself, and not only its derivatives. By using the fermions $\theta_I$ we avoid such terms.%
} %
We fix kappa gauge by imposing the condition
\begin{equation}\label{eq:kappa-gauge-fixing}
  \Gamma^+ \theta_I = 0, \qquad 
  \Gamma^{\pm} = \frac{1}{2} \bigl( \cos\varphi \Gamma^5 + \sin\varphi \Gamma^8 \pm \Gamma^0 \bigr) .
\end{equation}
By further introducing a different set of projectors
\begin{equation}
  \begin{aligned}
    \mathcal{P}_1 &= \frac{ 1 + \Gamma^{1234} }{2} \frac{ 1 + \Gamma^{1267} }{2} ,\qquad 
    \mathcal{P}_2 = \frac{ 1 + \Gamma^{1234} }{2} \frac{ 1 - \Gamma^{1267} }{2} , \\
    \mathcal{P}_3 &= \frac{ 1 - \Gamma^{1234} }{2} \frac{ 1 + \Gamma^{1267} }{2} ,\qquad
    \mathcal{P}_4 = \frac{ 1 - \Gamma^{1234} }{2} \frac{ 1 - \Gamma^{1267} }{2} ,
  \end{aligned}
\end{equation}
we can split the fermions into four groups
\begin{equation}
  \mathcal{P}_i \theta_I^{(i)} = \theta_I^{(i)} , \qquad i = 1,2,3,4.
\end{equation}
As we will see below, this divides the fermions according to mass of the fluctuation. After fixing kappa gauge, each of the eight spinors $\theta_I^{(i)}$ (for $i=1,2,3,4$ and $I=1,2$) contain a single complex fermionic degree of freedom. In the following we will therefore write out the action directly in terms of eight complex components $\theta_{Ii}$ and their complex conjugates $\bar{\theta}_{Ii}$. In appendix~\ref{app:spinors-and-grassmanns} explicit expressions for the 32-component spinors $\theta_I$ in terms of the components $\theta_{Ii}$.

To write down the gauge-fixed Lagrangian and supercurrents in a compact form we finally introduce the complex bosonic fields\footnote{%
  The leading order bosonic Lagrangian and supercurrents can further be compactly expressed in terms of the fields
  \begin{equation}
    W = w - i \psi , \qquad \bar{W} = w + i \psi .
  \end{equation}
  However, the compact $\algU(1)$ isometry acting on $W$ is broken at higher orders.
}%
\begin{equation}
  \begin{aligned}
    Z &= -z_2 + iz_1, \quad &
    Y &= -y_3 - iy_4, \quad &
    X &= -x_6 - ix_7, \\
    \bar{Z} &= -z_2 - iz_1, &
    \bar{Y} &= -y_3 + iy_4, &
    \bar{X} &= -x_6 + ix_7.
  \end{aligned}
\end{equation}
The quadratic-in-fermions terms in the gauge-fixed GS Lagrangian is then given by
\begin{align}
  \mathcal{L}_{\text{F}}^{(2)} =
  + & i\bar{\theta}_{11} \bigl( \dot{\theta}_{11} - i \tilde{q} \pri{\theta}_{21} + q \pri{\theta}_{11} \bigr)
  + i\bar{\theta}_{21} \bigl( \dot{\theta}_{21} + i \tilde{q} \pri{\theta}_{11} - q \pri{\theta}_{21} \bigr)
  \\ \nonumber
  + & i\bar{\theta}_{12} \bigl( \dot{\theta}_{12} - i \tilde{q} \pri{\theta}_{22} + q \pri{\theta}_{12} \bigr)
  + i\bar{\theta}_{22} \bigl( \dot{\theta}_{22} + i \tilde{q} \pri{\theta}_{12} - q \pri{\theta}_{22} \bigr)
  - \sin^2\varphi \bigl( \bar{\theta}_{12} \theta_{12} - \bar{\theta}_{22} \theta_{22} \bigr)
  \\ \nonumber
  + & i\bar{\theta}_{13} \bigl( \dot{\theta}_{13} - i \tilde{q} \pri{\theta}_{23} + q \pri{\theta}_{13} \bigr)
  + i\bar{\theta}_{23} \bigl( \dot{\theta}_{23} + i \tilde{q} \pri{\theta}_{13} - q \pri{\theta}_{23} \bigr)
  - \cos^2\varphi \bigl( \bar{\theta}_{13} \theta_{13} - \bar{\theta}_{23} \theta_{23} \bigr)
  \\ \nonumber
  + & i\bar{\theta}_{14} \bigl( \dot{\theta}_{14} - i \tilde{q} \pri{\theta}_{24} + q \pri{\theta}_{14} \bigr)
  + i\bar{\theta}_{24} \bigl( \dot{\theta}_{24} + i \tilde{q} \pri{\theta}_{14} - q \pri{\theta}_{24} \bigr)
  - \bigl( \bar{\theta}_{14} \theta_{14} - \bar{\theta}_{24} \theta_{24} \bigr) .
\end{align}
We note that the fermions $\theta_{I1}$, $\theta_{I2}$, $\theta_{I3}$ and $\theta_{I4}$ have mass $0$, $\sin^2\varphi$, $\cos^2\varphi$ and $1$, respectively. Furthermore, for the case of $q=1$ (and hence $\tilde{q}=0$) the fermions are all purely left- or right-moving on the worldsheet. The cubic order corrections to the fermionic Lagrangian can be found in appendix~\ref{app:cubic}.

\subsection{The off-shell symmetry algebra \texorpdfstring{$\algA$}{A}}
\label{sec:algebra-A}

The gauge-fixed action obtained at the end of the last subsection has four 
supersymmetries that commute with the Hamiltonian. In this subsection we write down the expressions for the associated supercurrents. We relax the level-matching condition and determine the algebra $\algA$ of the supercharges. We find that the off-shell (\ie, non-level matched) algebra $\algA$ contains four central elements $\genH$, $\genM$, $\gen{C}$ and $\overline{\gen{C}}$. We also determine the relationship between $\gen{C}$ and the worldsheet momentum $p_{\text{w.s.}}$. The resulting expressions are similar to those appearing in the off-shell symmetry algebra of $\AdS_3 \times \Sphere^3 \times \Torus^4$~\cite{Borsato:2014exa,Borsato:2014hja,Lloyd:2014bsa}.

\subsubsection{Supercurrents}
\label{sec:supercurrents}

After gauge fixing there are in total four conserved supercurrents. Below we will write expressions for the components of the two currents $j_{\sL}^{\mu}$ and $j_{\sR}^{\mu}$. The other two currents,  $\bar{\jmath}_{\sL}^{\,\mu}$ and $\bar{\jmath}_{\sR}^{\,\mu}$, can be obtained by complex conjugation. The labels ``L'' and ``R'' refer to chirality in the dual $\CFT_2$.

To quadratic order in the transverse fields, the $\tau$-components of the supercurrents are given by
\begin{equation} % Corresponds to J3 in the Mathematica file
  \begin{aligned}
    j_{\sL}^{\tau} = \tfrac{1}{2}e^{-i\pi/4} e^{+ix^-} \smash{\Bigl(}
    & + 2 P_{\bar{Z}} \theta_{14} + \pri{Z} ( i \tilde{q} \theta_{24} - q \theta_{14} ) + i Z \theta_{14}
    \\
    & - 2i P_{Y} \bar{\theta}_{13} - \pri{\bar{Y}} ( \tilde{q} \bar{\theta}_{23} - i q \bar{\theta}_{13} ) - \cos^2\varphi \, \bar{Y} \bar{\theta}_{13}
    \\
    & - 2i P_{X} \bar{\theta}_{12} - \pri{\bar{X}} ( \tilde{q} \bar{\theta}_{22} - i q \bar{\theta}_{12} ) - \sin^2\varphi \, \bar{X} \bar{\theta}_{12}
    \\
    & - i ( P_w + i P_{\psi} ) \bar{\theta}_{11} -  ( \pri{w} + i \pri{\psi} )  ( \tilde{q} \bar{\theta}_{21} - i q \bar{\theta}_{11} )
    \Bigr),
  \end{aligned}
\end{equation}%
and
\begin{equation} % Corresponds to J2 in the Mathematica file
  \begin{aligned}
    j_{\sR}^{\tau} = \tfrac{1}{2}e^{-i\pi/4} e^{+ix^-} \bigl(
    &+ 2 P_Z \bar{\theta}_{24} + \pri{\bar{Z}} ( i \tilde{q} \bar{\theta}_{14} + q \bar{\theta}_{24} ) + i \bar{Z} \bar{\theta}_{24}
    \\
    &- 2i P_{\bar{Y}} \theta_{23} - \pri{Y} ( \tilde{q} \theta_{13} + i q \theta_{23} ) - \cos^2\varphi \, Y \theta_{23}
    \\
    &- 2i P_{\bar{X}} \theta_{22} - \pri{X} ( \tilde{q} \theta_{12} + i q \theta_{22} ) - \sin^2\varphi \, X \theta_{22}
    \\
    &- i ( P_w - i P_{\psi} ) \theta_{21} -  ( \pri{w} - i \pri{\psi} )  ( \tilde{q} \theta_{11} + i q \theta_{21} )
    \bigr) .
  \end{aligned}
\end{equation}
The $\sigma$-components of the currents are given by
\begin{equation} % Corresponds to J3 in the Mathematica file
  \begin{aligned}
    j_{\sL}^{\sigma} = \tfrac{1}{2}e^{-i\pi/4} e^{+ix^-} \bigl(
    &- \pri{Z} \theta_{24} - ( 2 P_{\bar{Z}} + i Z ) ( i \tilde{q} \theta_{14} - q \theta_{24} )
    \\
    &+ i \pri{\bar{Y}} \bar{\theta}_{23} + ( 2 P_Y - i \cos^2\varphi \, \bar{Y} ) ( \tilde{q} \bar{\theta}_{13} - i q \bar{\theta}_{23} )
    \\
    &+ i \pri{\bar{X}} \bar{\theta}_{22} + ( 2 P_X - i \sin^2\varphi \, \bar{X} ) ( \tilde{q} \bar{\theta}_{12} - i q \bar{\theta}_{22} )
    \\
    &+ i ( \pri{w} + i \pri{\psi} ) \bar{\theta}_{23} + ( P_{w} + i P_{\psi} ) ( \tilde{q} \bar{\theta}_{11} - i q \bar{\theta}_{21} )
    \bigr).
  \end{aligned}
\end{equation}
and
\begin{equation} % Corresponds to J3 in the Mathematica file
  \begin{aligned}
    j_{\sR}^{\sigma} = \tfrac{1}{2}e^{-i\pi/4} e^{+ix^-} \bigl(
    &- \pri{\bar{Z}} \bar{\theta}_{24} - ( 2 P_Z + i \bar{Z} ) ( i \tilde{q} \bar{\theta}_{14} + q \bar{\theta}_{24} )
    \\
    &+ i \pri{Y} \theta_{23} + ( 2 P_{\bar{Y}} - i \cos^2\varphi \, Y ) ( \tilde{q} \theta_{13} + i q \theta_{23} )
    \\
    &+ i \pri{X} \theta_{22} + ( 2 P_{\bar{X}} - i \sin^2\varphi \, X ) ( \tilde{q} \theta_{12} + i q \theta_{22} )
    \\
    &+ i ( \pri{w} + i \pri{\psi} ) \theta_{23} + ( P_{w} - i P_{\psi} ) ( \tilde{q} \theta_{11} + i q \theta_{21} )
    \bigr).
  \end{aligned}
\end{equation}
The next-to-leading order in transverse bosons corrections to the currents are given in appendix~\ref{app:cubic}.
Using the attached Mathematica program, we have checked using the equations of motion derived from the Lagrangians presented in the previous sections that the above currents plus their higher order corrections satisfy the conservation equations $\partial_{\mu} j_I^{\mu}=0$ to cubic order in transverse bosons.
In the above expressions we have included a \emph{non-local} dependence on the non-dynamic field $x^-$. These exponential factors are essential when checking the current conservation at cubic order in transverse bosons. As we will see below, these terms are responsible for the central extension of the off-shell symmetry algebra.

\subsubsection{The algebra from the supercurrents}

The supercurrents presented above give rise to four supercharges
\begin{equation}
  \genQ_{\sL} = \int d\sigma j_{\sL}^{\tau} , \qquad
  \genQ_{\sR} = \int d\sigma j_{\sR}^{\tau} , \qquad
  \overline{\genQ}_{\sL} = \int d\sigma \bar{\jmath}_{\sL}^{\,\tau} , \qquad
  \overline{\genQ}_{\sR} = \int d\sigma \bar{\jmath}_{\sR}^{\,\tau} .
\end{equation}
We can find the algebra satisfied by these charges at a classical level by calculating Poisson brackets. To do this we first need to know the Poisson bracket of the fermions. 
To leading order these are given by~\footnote{The contributions arising from the Poisson bracket of two bosons need not be considered since they contribute to the algebra at next-to-leading order in fermions while our supercurrents are only valid up to leading order in fermions. As a result, the expression for the central charges presented in this section
involve only the bosonic fields.}
\begin{equation}
  \acommPB{\bar{\theta}_{Ii}}{\theta_{Jj}} = -i \delta_{IJ} \delta_{ij} \delta(x-y) , 
  \qquad
  \acommPB{\theta_{Ii}}{\theta_{Jj}} = 0 .
\end{equation}
These expression receive corrections that are quadratic in the transverse bosonic fields. We will not explicitly write out the corrections here. However, we have checked that the algebra presented below is preserved by the cubic-in-bosons currents, and in performing that calculation the corrections to the Poisson brackets of the fermions are essential.

Taking the Poisson bracket between a supercharge and its complex conjugate we find
\begin{equation}
  \label{eq:comm-rel-QL-QLb-and-QR-QRb}
  \begin{aligned}
    \acommPB{ \genQ_{\sL} }{ \overline{\genQ}_{\sL} } &= -\frac{i}{2} \bigl(\gen{H} + \gen{M}\bigr) , \\
    \acommPB{ \genQ_{\sR} }{ \overline{\genQ}_{\sR} } &= -\frac{i}{2} \bigl(\gen{H} - \gen{M}\bigr) ,
  \end{aligned}
\end{equation}
where the Hamiltonian density $\gen{H}$ is given to cubic order in transverse bosons by
\begin{align}
  \gen{H}  =&
  \tfrac{1}{2} \int d\sigma \smash{\Bigl(}
  p_z^2 + p_y^2 + p_x^2 + p_w^2 + p_{\psi}^2
  + \pri{z}^2 + \pri{y}^2 + \pri{x}^2 + \pri{w}^2 + \pri{\psi}^2
  \\ &\qquad \nonumber
  + z^2 + \cos^4\!\varphi \, y^2 + \sin^4\!\varphi \, x^2
  - 2q \epsilon^{ij} \bigl(
  z_i \pri{z}_j + \cos^2\!\varphi \, y_i \pri{y}_j + \sin^2\!\varphi \, x_i \pri{x}_j
  \bigr)
  \smash{\Bigr)}
  \\ & \nonumber
  -\tfrac{1}{2} \sin(2\varphi)
  \smash{\Bigl(}
  p_\psi \bigl( \cos^2\!\varphi \, y^2 - \sin^2\!\varphi \, x^2 \bigr)
  % \\ &\qquad\qquad
  - q \epsilon^{ij} \bigl( p_\psi ( y_i \pri{y}_j - x_i \pri{x}_j ) + \pri{\psi} ( p_{y^i} y_j - p_{x^i} x_j ) \bigr)
  \smash{\Bigr)} ,
\end{align}
and the charge $\gen{M}$ is given by\footnote{This expression for $\gen{M}$ is exact at least to quartic order. Moreover, $\gen{M}$ is a conserved quantity of the bosonic Hamiltonian to \emph{all} orders in the transverse fields.}
\begin{equation}
  \begin{aligned}
    \gen{M} = - \int d\sigma \smash{\Bigl(}
    &\epsilon^{ij} \bigl(
    p_{z^i} z_j + \cos^2\varphi \, p_{y^i} y_j + \sin^2\varphi \, p_{x^i} x_j
    \bigr)
    \\ &\quad
    + q \bigl( p_{z^i} \pri{z}^i + p_{y^i} \pri{y}^i + p_{x^i} \pri{x}^i + p_{w^i} \pri{w}^i + p_{\psi^i} \pri{\psi}^i \bigr) 
    \smash{\Bigr)}.
  \end{aligned}
\end{equation}
The second line gives a term proportional to the world sheet momentum $p_{\text{w.s.}}$. On shell, \ie, for $p_{\text{w.s.}} = 0$, the $\algU(1)$ charge $\gen{M}$ is given by a combination of angular momenta in $\AdS_3 \times \Sphere^3 \times \Sphere^3$, and the anti-commutation relations~\eqref{eq:comm-rel-QL-QLb-and-QR-QRb} are part of the $\algD{\alpha}^2$ superisometry algebra of the string background.

Let us now consider the Poisson bracket between $\genQ_{\sL}$ and $\genQ_{\sR}$. These supercharges belong to two different $\algD{\alpha}$ algebras and therefore anti-commute on shell. When we relax the level-matching condition we find
\begin{equation}
  \begin{aligned}
    \acommPB{ \genQ_{\sL} }{ \genQ_{\sR} }
%    = -i \gen{C}
    =
    \frac{\tilde{q}}{2} \int d\sigma \smash{\Bigl(} &
    \partial_{\sigma} \bigl( e^{2ix^-} \bigr) 
    + \tfrac{1}{2} \partial_{\sigma} \bigl( e^{2ix^-} ( z^2 - \cos^2\varphi \, y^2 - \sin^2\varphi \, x^2 ) \bigr)
    \\ &
    + \tfrac{1}{8} e^{2ix^-} \partial_{\sigma} ( z^2 - \cos^2\varphi \, y^2 - \sin^2\varphi \, x^2 )^2
    \smash{\Bigr)} .
  \end{aligned}
\end{equation}
The above expression is written out to \emph{quartic} order in bosons since it is quite compact even to this order. To obtain it, we used the cubic-in-bosons super-currents and corrected Poisson brackets contained in the Mathematica file attached to this paper. Using partial integration in the second line we obtain one term that integrates to zero as well as a term that is higher order in transverse fields. Similarly, the second term in the first line vanishes upon integration. The remaining integral is non-vanishing since the field $x^-$ is non-trivial at $\sigma \to \pm \infty$. Hence we are left with a non-trivial Poisson bracket
\begin{equation}\label{eq:commu-rel-QL-QR}
  \acommPB{ \genQ_{\sL} }{ \genQ_{\sR} }
  = -i \gen{C} ,
\end{equation}
where the central charge $\gen{C}$ evaluates to
\begin{equation}
  \label{eq:result-central-charge-C}
  \gen{C}
  =
  \frac{i\zeta\tilde{q}}{2} \bigl( e^{i p_{\text{w.s.}}} - 1 \bigr) .
\end{equation}
The constant $\zeta$ is given by $\zeta = \exp(2ix^-(-\infty))$. Since a physical state satisfies $p_{\text{w.s.}} \in 2\pi\Integers$, the charge $\gen{C}$ vanishes when acting on such a state, as expected.

The Poisson bracket between supercharges $\overline{\genQ}_{\sL}$ and $\overline{\genQ}_{\sR}$ can be obtained from equation~\eqref{eq:commu-rel-QL-QR} by complex conjugation\footnote{%
  Note that in our conventions the Poisson bracket of two Grassmann odd quantities is anti-Hermitian.%
}%
\begin{equation}\label{eq:commu-rel-QLb-QRb}
  \acommPB{ \overline{\genQ}_{\sL} }{ \overline{\genQ}_{\sR} }
  = -i \overline{\gen{C}} .
\end{equation}

In summary, we have investigated the symmetry algebra of the gauge-fixed type IIB string theory on $\AdS_3\times\Sphere^3\times \Sphere^3\times S^1$. On shell this algebra is given by
\begin{equation}
  \algSU(1|1)^2 \subset \algD{\alpha}^2 .
\end{equation}
Going off shell, by letting the world sheet momentum take arbitrary values, we showed that this algebra is enlarged by two additional central charges $\gen{C}$ and $\overline{\gen{C}}$. We denote the resulting algebra by
\begin{equation}
  \mathcal{A} = \algPSU(1|1)^2_{\ce} .
\end{equation}

\section{Representations of $\mathcal{A}$ at quadratic order in fields}
\label{sec:quadraticrepr}

In this section we will present the short representations of the symmetry algebra at quadratic order in the fields.

\subsection{Off-shell symmetry algebra}

The $\AdS_3\times \Sphere^3\times \Sphere^3\times\Sphere^1 $ background preserves four supercharges after light-cone gauge fixing. This is half of the amount preserved by the $\AdS_3\times \Sphere^3\times \Torus^4$ background, which can be seen as a limit of the case at hand when $\alpha\to 0$ or $\alpha\to 1$.
In the previous section we introduced four supercharges
\begin{equation}\label{eq:relation-supercharges-supercurrents}
  \genQ_{\sL} = \int d\sigma j_{\sL}^{\tau} , \qquad
  \genQ_{\sR} = \int d\sigma j_{\sR}^{\tau} , \qquad
  \overline{\genQ}_{\sL} = \int d\sigma \bar{\jmath}_{\sL}^{\,\tau} , \qquad
  \overline{\genQ}_{\sR} = \int d\sigma \bar{\jmath}_{\sR}^{\,\tau} .
\end{equation}
As we found there, these charges satisfy the centrally extended $\algPSU(1|1)^2$ algebra\footnote{
  In the rest of the paper we will write the algebra in terms of canonical anti-commutators instead of Poisson brackets. The two notations are related by
  \begin{equation*}
    \acomm{A}{B} = i \acommPB{A}{B} .
  \end{equation*}
}
\begin{equation}
  \begin{aligned}
    \{ \QL , \QbL \} &= \frac{1}{2} (\gen{H}+ \gen{M}) ,
    \qquad
    &
    \{ \QL , \QR \} &= \gen{C}\, ,
    \\
    \{ \QR , \QbR \} &= \frac{1}{2} (\gen{H} - \gen{M}) ,
    \qquad
    &
    \{ \QbL , \QbR \} &= \overline{\gen{C}} ,
  \end{aligned}
\end{equation}
where $\gen{H}$ is the Hamiltonian, $\gen{M}$ is an angular momentum on shell and $\gen{C}$ and $\overline{\gen{C}}$ are central charges appearing off-shell~\cite{Beisert:2005tm,Arutyunov:2006ak}.

\subsection{Irreducible representations}

To make the representations of this symmetry algebra more transparent it is  convenient to rewrite the charges in terms of oscillators, which is straightforward at quadratic order in the fields. To this end, let us introduce the wave-function parameters $f_{\sL,\sR}, g_{\sL,\sR}$ and the dispersion relations~$\omega_{\sL,\sR}$,
\begin{equation}
  \label{eq:wavefun}
  \begin{aligned}
    g_{\sL} (p,m_j) &= -\frac{\tilde{q} \, p}{2f_{\sL}(p,m_j)},
    \quad &
    g_{\sR} (p,m_j) &= -\frac{\tilde{q} \, p}{2f_{\sR}(p,m_j)},
    \\
    f_{\sL}(p,m_j) &= \sqrt{\frac{|m_j|+ q \, p +\omega_{\sL}(p,m_j)}{2}},
    \quad &
    f_{\sR}(p,m_j) &= \sqrt{\frac{|m_j|- q \, p +\omega_{\sR}(p,m_j)}{2}},
    \\
    \omega_{\sL}(p,m_j) &= \sqrt{p^2 + 2 \, |m_j|\, q\, p + m_j^2},
    \quad &
    \omega_{\sR}(p,m_j) &= \sqrt{p^2 - 2 \, |m_j|\, q\, p + m_j^2},
  \end{aligned}
\end{equation}
with the labels L,R standing for ``left'' and ``right''.%
\footnote{%
  This corresponds to left and right chirality in the dual $\CFT_{2}$.
}
All  these parameters depend on the momentum $p$, on the NS-NS flux coefficient~$q$, and on the oscillators' mass~$|m_j|$. We expect $|m_j|$ to take values 1, $\alpha$, $1-\alpha$ and~0 for the bosonic oscillators corresponding to modes on $\AdS_3$, on each of the two spheres, and to the flat coordinates, respectively (and similarly for their fermionic partners). Hence, we expect to find four representations of the symmetry algebra (one for each mass), which may be further reducible.
%The functions $f$ and $g$ are related to the frequency $\omega$ as
%\begin{equation}
%f_{\sL}(p,\mu)^2 + g_{\sL}(p,\mu)^2 =\omega_{\sL}(p,\mu),
%\qquad
%f_{\sR}(p,\mu)^2 + g_{\sR}(p,\mu)^2 =\omega_{\sR}(p,\mu).
%\end{equation}
We can then schematically write the bosons in terms of creation and annihilation operators as usual,
\begin{equation}
  \begin{aligned}
    X &\approx  \int dp \left( \frac{1 }{\sqrt{\omega_{\sL}}} \, a_{\sL}^\dagger(p)\ e^{-i\, p\sigma} + \frac{1}{\sqrt{\omega_{\sR}}}\, a_{\sR}(p)  \ e^{i\, p\sigma} \right),
    \\
    P &\approx i\,\int dp \left( \sqrt{\omega_{\sL}}\, a_{\sL}^\dagger(p)  \ e^{-i\, p\sigma} - \sqrt{\omega_{\sR}}\, a_{\sR}(p)  \ e^{i\, p\sigma} \right),
  \end{aligned}
\end{equation}
and similarly for the fermions
\begin{equation}
  \begin{aligned}
    \theta^{\sL} &\approx  \int dp \left( \frac{g_{\sR}}{\sqrt{\omega_{\sR}}} \, d_{\sR}^\dagger \ e^{-i\, p\sigma} -
      \frac{f_{\sL}}{\sqrt{\omega_{\sL}}} \, d_{\sL} \ e^{i\, p\sigma} \right),
    \\
    \theta^{\sR} &\approx \int dp \left( \frac{g_{\sL}}{\sqrt{\omega_{\sL}}} \, d_{\sL}^\dagger \ e^{-i\, p\sigma} -
      \frac{f_{\sR}}{\sqrt{\omega_{\sR}}} \, d_{\sR} \ e^{i\, p\sigma} \right).
  \end{aligned}
\end{equation}
Note that like in reference~\cite{Lloyd:2014bsa} we have to introduce ``left'' and ``right'' oscillators with appropriate wave-function parameters due to the presence of the parity-breaking NS-NS flux, \ie, since $q\neq 0$. Moreover, for each value of the mass there will be one set of oscillators
\begin{equation}
a_{\sL\,j}, a_{\sR\,j} \text{ and } d_{\sL\,j}, d_{\sR\,j} \qquad \text{ with } j\in\{1,2,3,4\} ,
\end{equation}
for a total of $8+8$ bosonic and fermionic oscillators, whose precise definition can be found in appendix~\ref{app:charges}.

In terms of these oscillators, the supercharges take a simple form:
\begin{equation}
  \label{eq:supercharges}
  \begin{aligned}
    \QL = \int dp \smash{\,\Bigl(\!}
    &-f_{\sL}(p,1) \ a_{\sL \, 4}^\dagger(p) d_{\sL\, 4}(p) - g_{\sR}(p,1) \ d_{\sR \, 4}^\dagger(p) a_{\sR\, 4}(p) \\
    &+ \sum_{j=1}^3 \bigl( f_{\sL}(p,m_j) \ d_{\sL \, j}^\dagger(p) a_{\sL\, j}(p) + g_{\sR}(p,m_j) \ a_{\sR \, j}^\dagger(p) d_{\sR\, j}(p) \bigr) 
    \smash{\!\Bigr)},
    \\
    \QR = \int dp \smash{\,\Bigl(\!}
    &-f_{\sR}(p,1) \ a_{\sR \, 4}^\dagger(p) d_{\sR\, 4}(p) - g_{\sL}(p,1) \ d_{\sL \, 4}^\dagger(p) a_{\sL\, 4}(p) \\
    &+ \sum_{j=1}^3 \bigl( f_{\sR}(p,m_j) \ d_{\sR \, j}^\dagger(p) a_{\sR\, j}(p) + g_{\sL}(p,m_j) \ a_{\sL \, j}^\dagger(p) d_{\sL\, j}(p) \bigr) 
    \smash{\!\Bigr)}.
  \end{aligned}
\end{equation}
On the first line of each equation\footnote{The minus sign appearing for the contribution of $|m|=1$ could be reabsorbed \eg by redefining the fermionic fields of mass $|m|=1$, or the map that relates them to the corresponding creation and annihilation operators. We prefer the present convention, so that other expressions are more natural.} we wrote the contribution of the oscillators with mass $|m|=1$. All the remaining ones can be grouped together, as the representations for masses $|m|=0,1-\alpha,\, \alpha$ have the same  grading. We read off two irreducible representations for each mass, each labelled by left or right,%
\footnote{%
As we will see in the next subsection, this label is not entirely appropriate for the massless representations.
}
 for a total of eight two-dimensional irreducible representations. They are all short representations, satisfying the shortening condition
\begin{equation}
\label{eq:shortening}
\gen{H}^2=\gen{M}^2+4\gen{C}\overline{\gen{C}}\,.
\end{equation}
The eigenvalues of the central charges $\gen{M}$ and $\gen{H}$ on each module are, at this order in the field expansions~\footnote{The corresponding dispersion relations for the massive modes of the mixed flux $\AdS_3\times \Sphere^3\times \Torus^4$ were first discussed in~\cite{Hoare:2013pma,Hoare:2013ida,Hoare:2013lja}.}
\begin{equation}
\begin{aligned}
\label{eq:quadraticcentrcharges}
\gen{M} &= m + qp & =&\left\{
\begin{array}{cc}
\phantom{-} |m| + qp & \text{left},
\\
- |m| + qp & \text{right},
\end{array}
\right.
\\
\gen{H} &= \sqrt{p^2 + 2 \, m\, q\, p + m^2} & =&\left\{
\begin{array}{cc}
\sqrt{p^2 + 2 \, |m|\, q\, p + m^2} & \text{left},
\\
\sqrt{p^2 - 2 \, |m|\, q\, p + m^2} & \text{right}.
\end{array}
\right.
\end{aligned}
\end{equation}
with $|m|=1,\,1-\alpha,\,\alpha,0$. Consistently, the off-shell central charges are $\gen{C}=\overline{\gen{C}}=-\tfrac{1}{2}\tilde{q}\,\gen{P}$ for all representations at quadratic order in the fields.

Finally, it is interesting to note that equation~\eqref{eq:supercharges} possesses a discrete symmetry under swapping ``L'' and ``R'' labels everywhere. This is just a generalisation of the \emph{left-right symmetry} (LR symmetry) introduced in reference~\cite{Borsato:2012ud}.

\subsection{Heavy representations}
\label{sec:heavymode}

There are two heavy representations with $|m|=1$. These modes look similar
to the heavy modes of~$\AdS_4\times\CP^3$ superstrings,\footnote{%
  To see this similarity consider the case $\alpha = 1/2$. The supergroup $\grpD{1/2}$ is the same as $\grpOSp(4|2)$. In the coset sigma model the heavy modes are embedded in $\grpOSp(4|2)^2$ in essentially the same way the heavy modes of $\AdS_4 \times \CP^3$ are embedded in $\grpOSp(6|4)$.%
} %
which in fact are \emph{composite}.~\cite{Zarembo:2009au} This means that in that theory the heavy modes should not be regarded as part of the asymptotic particle spectrum, but should instead be understood as a compound of two real lighter particles. Therefore, in the Bethe ansatz description of the spectrum there are no momentum-carrying nodes corresponding to the heavy modes---instead, these are represented by stacks of two (lighter) Bethe roots.%
\footnote{See reference~\cite{Klose:2010ki} for a review of $\AdS_4\times\CP^3$ integrability and for an extensive list of references on the subject.}

It is natural to wonder whether something similar may happen here. At the order in the near-plane-wave expansion that we are considering, this is certainly allowed kinematically. Let us consider two particles of mass $m_1=\alpha$, $m_2=1-\alpha$ and momenta $p_1=\alpha\,p$, $p_2=(1-\alpha)p$. Then their total energy is
\begin{equation}
E_{\text{tot}}=\sqrt{p_1^2 + 2 \, m_1\, q\, p_1 + m_1^2}+\sqrt{p_2^2 + 2 \, m_2\, q\, p_2 +m_2^2}=\sqrt{p^2 + 2 \, q\, p + 1} ,
\end{equation} 
with the total mass adding up to~$1$ and the total momentum adding up to~$p$. At higher orders in perturbation theory, as we will see in the next section, the central charges and hence the dispersion relations get deformed. If the heavy modes are indeed composite, we would expect that their representations reveal themselves as long (\ie, they are ``accidentally short'' at this order), fusing up with some other multi-particle excitation of mass one---note that all long representations are four-dimensional. If all this happened, we would not need to consider the heavy modes in our integrable S~matrix and Bethe ansatz.

A different possibility is that the heavy modes do transform in genuinely short representations, but are \emph{bound states} of lighter excitations. This would mean that we can get the heavy modes by tensoring two light representations. The constituents should have  suitable (complex) momenta, so that the resulting four-dimensional representation may be reducible yielding a short representation by a quotient, much like in reference~\cite{Arutyunov:2008zt}.%
\footnote{%
Here all short representations, including the bound-state one, would be two-dimensional. In the case of the $\psu(2|2)$ representations for $\AdS_5\times\Sphere^5$ superstrings, the short representation of bound-state number~$M$ has dimension~$4M$.
} 
There is an interesting difference between the would-be bound state with $|m|=1$ and the ones familiar from $\AdS_5\times\Sphere^5$. The latter can be geometrically interpreted as bound states of giant magnons on the sphere~\cite{Dorey:2006dq}, and correspond to the totally symmetric combination of the constituent representations. Such magnon bound states also exist here, and can be constructed out of two light excitations of \emph{the same mass}. The heavy mode instead sits in a representation with opposite grading than its constituents, and as such must come from the anti-symmetric combination%
\footnote{%
For the sake of this argument we can take $\alpha=1/2$.
}
of its constituents. While we will not investigate this further here, it is interesting to note that this would yield a different kinematic condition for bound states---much like the one of the $\AdS_5\times\Sphere^5$ \emph{mirror} theory~\cite{Arutyunov:2007tc}. 
In any case, should the heavy modes be bound states, we could  consistently consider scattering processes involving them. However their scattering matrices would be uniquely fixed, under the assumption of integrability, 
in terms of the ones of light modes through \emph{fusion}~\cite{Kulish:1981gi}.

While there is some evidence that these heavy modes may be indeed composite~\cite{Sundin:2012gc}, we will not assume this for the time being. We will proceed by treating them as good asymptotic states, and write down restrictions on their scattering matrices. Such S~matrices may either not exists (``composite'' scenario) or be redundant (``bound state'' scenario).

\subsection{Massless representations}
\label{sec:massless-reps}

Looking at equation~\eqref{eq:quadraticcentrcharges} we see that at $m=0$ ``left'' and ``right'' representations have the same central charges. In this sense such a distinction is arbitrary, and indeed one can check that a massless ``left'' representation is isomorphic to a ``right'' one of opposite grading. Like in reference~\cite{Lloyd:2014bsa} the change-of-basis matrix depends on the momentum through
\begin{equation}
\frac{a_{\sL}(p)}{b_{\sR}(p)}=-\text{sgn}\big(\sin\frac{p}{2}\big).
\end{equation}
Still, unlike what happens in the case of $\AdS_3\times\Sphere^3\times\Torus^4$, where the massless modules are rotated into each other by an additional $\su(2)$ symmetry, here the two massless representations are completely distinct. For the bosons this can be expected from the geometry, where the massless directions $w$ and $\psi$ correspond to coordinates of $\Sphere^1$ and $\Sphere^3 \times \Sphere^3$, respectively.

\subsection{The \texorpdfstring{$\alpha\to 1$}{alpha to 1} limit}

As we mentioned, when sending $\alpha\to1$ or $\alpha\to0$ we expect the off-shell symmetry algebra to have twice as many supercharges. This is the case for $\AdS_3\times\Sphere^3\times\Torus^4$, which is the background that we would obtain from $\AdS_3\times\Sphere^3\times\Sphere^3\times\Sphere^1$ in those limits, up to compactifying the flat directions. This symmetry enhancement would also require the short (two-dimensional) representations of $\psu(1|1)^2_{\ce}$ to join up into short (four-dimensional) representations of $\psu(1|1)^4_{\ce}$. Such a merging may be subtle in the quantum theory, especially if the heavy modes are indeed composite for generic values of~$0<\alpha<1$, see also reference~\cite{Sundin:2012gc}. Still, it is worth briefly examining whether there is any obstruction from a representation-theoretical point of view.

When sending $\alpha\to1$ we find that the two-dimensional representations of $\psu(1|1)^2_{\ce}$ do come in pairs: two with $m=+1$, two with $m=-1$ and four with~$m=0$. The two $m=+1$ representations have opposite grading, so that one of the $\psu(1|1)^4_{\ce}$ supercharges that are not in $\psu(1|1)^2_{\ce}$ can act on the bosonic $\psu(1|1)^2_{\ce}$ highest weight state $a^\dagger_{\sL\,3}\ket{0}$ at $\alpha=1$,  to give the fermionic one $d^\dagger_{\sL\,4}\ket{0}$. Things go similarly for $m=-1$ and for the massless modes. In that case, we have, \eg, that the representation of bosonic highest weight state $a^\dagger_{\sL\,2}\ket{0}$ becomes related to the one of fermionic highest weight state~$d^\dagger_{\sR\,1}\ket{0}$. It may appear unnatural to mix left and right representations. However, as discussed, the left and right \emph{massless} representations are isomorphic to the transpose of each other. Hence, there is no inconsistency. 

As the situation is perfectly symmetric when~$\alpha\to0$, we can conclude that the representations which we found are compatible with the symmetry enhancement to $\psu(1|1)^4_{\ce}$.
The matching of the $\psu(1|1)^2_{\ce}$ and $\psu(1|1)^4_{\ce}$ supercharges in the $\alpha \to 1$ limit is further described in appendix~\ref{app:a-to-1-limit}.

\section{Exact representations}
\label{sec:exact-rep}
From the analysis of the supercurrents in section~\ref{sec:algebra-A} we expect the off-shell symmetry algebra to be still given by $\mathcal{A}$ even at higher orders in the field expansion. However, the representations will be deformed with respect to the ones described above, which can be seen by looking at the central charges. In this section we present the exact short representations of $\mathcal{A}$.

\subsection{Central charges}
From equation~\eqref{eq:result-central-charge-C} we expect the central charges $\gen{C},\overline{\gen{C}}$ to take the form
\begin{equation}
  \label{eq:Cexact}
  \gen{C} = +\frac{i}{2}\,h(\lambda,q,\alpha) \left( e^{+i\gen{P}}-1 \right),
  \qquad
  \overline{\gen{C}} = -\frac{i}{2}\,h(\lambda,q,\alpha) \left( e^{-i\gen{P}}-1 \right),
\end{equation}
where $\gen{P}$ is the worldsheet momentum and the effective coupling $h(\lambda,q,\alpha)$ is related to the string tension by
\begin{equation}
  h(\lambda,q,\alpha)\approx\frac{\tilde{q}\,\sqrt{\lambda}}{2\pi},
\end{equation}
up to sub-leading orders in~$\sqrt{\lambda}$. It is worth noticing that here we expect such subleading contributions to depend on the geometrical parameter~$\alpha$~\cite{Abbott:2012dd,Beccaria:2012kb}. With this in mind, from now on for simplicity we will write  $h\equiv h(\lambda,q,\alpha)$.

In the presence of the NS-NS flux the charge $\gen{M}$ also depends on the momentum as
\begin{equation}
  \label{eq:Mexact}
  \gen{M} = m + \k \gen{P} ,
\end{equation}
with $m=+|m|$ on left representations, and $m=-|m|$ on right representations. The constant $\k$ is given by
\begin{equation}
  \k=\frac{q\sqrt{\lambda}}{2\pi} .
\end{equation}
Since $k = q\sqrt{\lambda}$ is the integer-valued coupling of the WZ term in the bosonic action, we expect $\k$ to be exact to all orders in $\sqrt{\lambda}$.

From the shortening condition~\eqref{eq:shortening} we find the dispersion relation
\begin{equation}
E_p = \sqrt{(m+ \k p )^2 + 4 h^2 \sin^2 \frac{p}{2}} .
\end{equation}
We now want to modify the representations introduced in section~\ref{sec:quadraticrepr} in such a way as to reproduce these central charges.

\subsection{Short representations}
\label{sec:shortreprs}

Let us now describe the most general short representations of $\algPSU(1|1)^2_{\ce}$. We introduce the coefficients $a,b$ and their complex conjugates $\bar{a},\bar{b}$, which will depend on the particle's momentum~$p$ and mass~$|m|$, as well as on~$h,\alpha$ and $q$. At zero momentum the symmetry algebra reduces to $\algSU(1|1)^2$. In the representations below we require the coefficients $b$ to vanish for $p=0$. Note that each representation in that case transforms under just one of the two $\algSU(1|1)$ algebras, as indicated by the labels $L$ and $R$.

As in~\cite{Borsato:2012ud} we can define a left module~$\varrho_{\sL}$ consisting of a boson $\phi^{\sL}$ and a fermion $\psi^{\sL}$
\begin{equation}
\boxed{\varrho_{\sL}:} \qquad\qquad
  \begin{aligned}
    \gen{Q}_{\smallL} \ket{\phi^{\sL}} &= a^{\sL} \ket{\psi^{\sL}} , \qquad &
    \overline{\gen{Q}}{}_{\smallL} \ket{\psi^{\sL}} &= \bar{a}^{\sL}_p \ket{\phi^{\sL}} , \\
    \overline{\gen{Q}}{}_{\smallR} \ket{\phi^{\sL}} &= \bar{b}^{\sL}_p \ket{\psi^{\sL}} , \qquad &
    \gen{Q}_{\smallR} \ket{\psi^{\sL}} &= b^{\sL} \ket{\phi^{\sL}} ,
  \end{aligned}
\end{equation}
where we decorated the representation parameters to remind ourselves that they pertain to the left module. Similarly, we have a right representation $\varrho_{\sR}$
\begin{equation}
\boxed{\varrho_{\sR}:} \qquad\qquad
  \begin{aligned}
    \gen{Q}_{\sR} \ket{\phi^{\sR}} &= a^{\sR} \ket{\psi^{\sR }} , \qquad &
    \overline{\gen{Q}}{}_{\sR} \ket{\psi^{\sR }} &= \bar{a}^{\sR}_p \ket{\phi^{\sR}} , \\
    \overline{\gen{Q}}{}_{\sL} \ket{\phi^{\sR}} &= \bar{b}^{\sR} \ket{\psi^{\sR}_p} , \qquad &
    \gen{Q}_{\sL} \ket{\psi^{\sR}} &= b^{\sR} \ket{\phi^{\sR}}, 
  \end{aligned}
\end{equation}
that is formally obtained from the previous one by exchanging the labels L and R on the supercharges, the states and the momentum-dependent coefficients.

We can obtain two more representations by changing the grading of the ones above, in other words by swapping the role of the boson and the fermion. Denoting these representations with a tilde, we find
\begin{equation}
\boxed{\widetilde{\varrho}_{\sL}:} \qquad\qquad
  \begin{aligned}
    \gen{Q}_{\smallL} \ket{\tilde{\psi}^{\sL}} &= a^{\sL} \ket{\tilde{\phi}^{\sL}} , \qquad &
    \overline{\gen{Q}}{}_{\smallL} \ket{\tilde{\phi}^{\sL}} &= \bar{a}^{\sL} \ket{\tilde{\psi}^{\sL }} , \\
    \overline{\gen{Q}}{}_{\smallR} \ket{\tilde{\psi}^{\sL }_p} &= \bar{b}^{\sL} \ket{\tilde{\phi}^{\sL}} , \qquad &
    \gen{Q}_{\smallR} \ket{\tilde{\phi}^{\sL}} &= b^{\sL} \ket{\tilde{\psi}^{\sL}} , 
  \end{aligned}
\end{equation}
and
\begin{equation}
\boxed{\widetilde{\varrho}_{\sR}:} \qquad\qquad
  \begin{aligned}
    \gen{Q}_{\smallR} \ket{\tilde{\psi}^{\sR}} &= a^{\sR} \ket{\tilde{\phi}^{\sR}} , \qquad &
    \overline{\gen{Q}}{}_{\smallR} \ket{\tilde{\phi}^{\sR}} &= \bar{a}^{\sR} \ket{\tilde{\psi}^{\sR}} , \\
    \overline{\gen{Q}}{}_{\smallL} \ket{\tilde{\psi}^{\sR}} &= \bar{b}^{\sR} \ket{\tilde{\phi}^{\sR}} , \qquad &
    \gen{Q}_{\smallL} \ket{\tilde{\phi}^{\sR}} &= b^{\sR} \ket{\tilde{\psi}^{\sR}} .
  \end{aligned}
\end{equation}

The representations so constructed automatically satisfy~\eqref{eq:shortening}. On the left representations we have
\begin{equation}
  \gen{H}=\big(|a^{\mathrlap{\sL}\phantom{\sR}}|^2+|b^{\mathrlap{\sL}\phantom{\sR}}|^2\big)\gen{1} ,
  \qquad
  \gen{M}=+\big(|a^{\mathrlap{\sL}\phantom{\sR}}|^2-|b^{\mathrlap{\sL}\phantom{\sR}}|^2\big)\gen{1} ,
  \qquad
  \gen{C}=a^{\mathrlap{\sL}\phantom{\sR}}b^{\mathrlap{\sL}\phantom{\sR}}\,\gen{1} ,
\end{equation}
while on the right representations we have
\begin{equation}
  \gen{H}=\big(|a^{\sR}|^2+|b^{\sR}|^2\big)\gen{1} ,
  \qquad
  \gen{M}=-\big(|a^{\sR}|^2-|b^{\sR}|^2\big)\gen{1} ,
  \qquad
  \gen{C}=a^{\sR}b^{\sR}\,\gen{1} .
\end{equation}

\subsection{Exact representation coefficients}
It is convenient to  parametrise the representation coefficients $a^{\sL},b^{\sL},a^{\sR},b^{\sR}$ and their complex conjugates by introducing the Zhukovski variables $x^\pm_{\sL\,p}$ and $x^\pm_{\sR\,p}$ that satisfy the constraints
\begin{equation}
  \label{eq:zhukovski2}
  \begin{gathered}
    \frac{x^+_{\sL\,p}}{x^-_{\sL\,p}}=e^{ip},
    \qquad
    x^+_{\sL\,p} +\frac{1}{x^+_{\sL\,p}} -x^-_{\sL\,p} -\frac{1}{x^-_{\sL\,p}} = \frac{2i \, (|m|+\k \, p)}{\h},\\
    \frac{x^+_{\sR\,p}}{x^-_{\sR\,p}}=e^{ip},
    \qquad
    x^+_{\sR\,p} +\frac{1}{x^+_{\sR\,p}} -x^-_{\sR\,p} -\frac{1}{x^-_{\sR\,p}} = \frac{2i \, (|m|-\k\, p)}{\h}.
  \end{gathered}
\end{equation}
These equations can be solved by setting
\begin{equation}
  \begin{gathered}
    x^{\pm}_{\sL\,p}=\frac{(|m| + \k p)+\sqrt{(|m| + \k p)^2 + 4\h^2 \sin^2(\frac{p}{2})}}{2\h\sin(\frac{p}{2})}e^{\pm\frac{i}{2}p} ,\\
    x^{\pm}_{\sR\,p}=\frac{(|m| - \k p)+\sqrt{(|m| - \k p)^2 + 4\h^2 \sin^2(\frac{p}{2})}}{2\h\sin(\frac{p}{2})}e^{\pm\frac{i}{2}p} .
  \end{gathered}
\end{equation}
Then we take the representation coefficients to be
\begin{equation}\label{eq:abparam}
  \begin{aligned}
    a^{\sL} &= \eta_p^{\sL}\, e^{i\xi},
    &\quad
    \bar{a}^{\sL} &= \eta_p^{\sL}\, e^{-ip/2} e^{-i\xi},
    &\quad
    b^{\sL} &= -\frac{\eta_p^{\sL}}{x^-_{\sL\, p}} e^{-ip/2} e^{i\xi},
    &\quad
    \bar{b}^{\sL} &= -\frac{\eta^{\sL}_p}{x^+_{\sL\,p}} e^{-i\xi},\\
    a^{\sR} &= \eta_p^{\sR}\, e^{i\xi},
    &\quad
    \bar{a}^{\sR} &= \eta_p^{\sR}\, e^{-ip/2} e^{-i\xi},
    &\quad
    b^{\sR} &= -\frac{\eta_p^{\sR}}{x^-_{\sR\, p}} e^{-ip/2} e^{i\xi},
    &\quad
    \bar{b}^{\sR} &= -\frac{\eta^{\sR}_p}{x^+_{\sR\,p}} e^{-i\xi},
  \end{aligned}
\end{equation}
with
\begin{equation}
  \label{eq:etadef}
  \eta_p^{\sL} = e^{ip/4}\sqrt{\frac{i\h}{2}(x^-_{\sL\, p} - x^+_{\sL\, p})} ,
  \qquad
  \eta_p^{\sR} = e^{ip/4}\sqrt{\frac{i\h}{2}(x^-_{\sR\, p} - x^+_{\sR\, p})}.
\end{equation}
Note that we have introduced an additional parameter $\xi$. This has to be set to zero for the one-particle representation to match the central charges~\eqref{eq:Cexact}--\eqref{eq:Mexact}. However, $\xi$ is needed to consistently define multi-particle representations~\cite{Arutyunov:2006yd}. In fact, if we consider a two-particle state with momenta $p_1, p_2$ and parameters $\xi_1,\xi_2$ and we require the central charges $\gen{C},\overline{\gen{C}}$ to match equation~\eqref{eq:Cexact}, we must make the non-local assignment~\cite{Arutyunov:2009ga,Sfondrini:2014via}
\begin{equation}
  \xi_1 = 0,
  \qquad\qquad
  \xi_2 = p_1/2.
\end{equation}
This amounts to defining a non-local coproduct for the off-shell symmetry algebra~\cite{Plefka:2006ze}.

In the previous section we saw that to leading order all excitations transform in short representations of the symmetry algebra $\mathcal{A}$. Assuming the representation remain short also at higher orders---as discussed in section~\ref{sec:heavymode} this is quite subtle for the heaviest modes---we can use the exact representations constructed above to organise the spectrum of world sheet excitations. This lead to eight exact representations, which can be grouped by mass and chirality as
\begin{center}
  \begin{tabular}{ccccc}
    \toprule
    & $|m|=1$ & $|m|=\alpha$ & $|m|=1-\alpha$ & $|m|=0$ \\
    \midrule
    L & $\widetilde{\varrho}_{\sL}$ & $\varrho_{\sL}$ & $\varrho_{\sL}$ & $\varrho_{\sL}\cong\widetilde{\varrho}_{\sR}$ \\
    R & $\widetilde{\varrho}_{\sR}$ & $\varrho_{\sR}$ & $\varrho_{\sR}$ & $\varrho_{\sR}\cong\widetilde{\varrho}_{\sL}$ \\
    \bottomrule
  \end{tabular}
\end{center}

\section{The integrable S~matrix}
\label{sec:int-s-matrix}

In this section we present the two-body world sheet S~matrix.
The off-shell symmetry algebra $\mathcal{A}$ severely restricts the form of the S-matrix. Integrability further limits the allowed scattering processes. 
Finally, we will require (braiding and physical) unitarity and crossing invariance, and use this to constrain the dressing phases appearing in the S~matrix.

\subsection{Allowed processes}

In an integrable theory the presence of higher conserved charges imposes strong constraints on which two-particle scattering processes can appear~\cite{Zamolodchikov:1978xm}.
Let us consider the scattering of two particles with quantum numbers $(p_1,m_1)$ and $(p_2,m_2)$, resulting in two particles $(p_1',m_1')$ and $(p_2',m_2')$. The central charges impose constraints on $(p_j',m_j')$, yielding
\begin{equation}
m_1+m_2=m_1'+m_2' , \qquad
p_1+p_2=p_1'+p_2' , \qquad
E_1+E_2=E_1'+E_2' ,
\end{equation}
where we also imposed invariance under worldsheet translations. This allows for a plethora of scattering channels. For instance if the masses $|m_1|,|m_2|$ take values $\alpha,1-\alpha$, the outgoing particles may have masses $|m_1'|, |m_2'|$ equal to
\begin{equation}
  \label{eq:badprocesses}
  \alpha, \, 1-\alpha, \qquad
  1-\alpha, \, \alpha, \qquad
  0, \, 1, \quad
  \text{or}\quad
  1,\, 0.
\end{equation}
In general the dependence of the outgoing momenta on the incoming ones is complicated. However, only one of the outcomes is compatible with integrability. If we require the conservation of higher charges of the form~\cite{Beisert:2004hm,Arutyunov:2004vx}
\begin{equation}
  \mathcal{Q}_n=\frac{i}{n-1}\left(\frac{1}{(x^{+}_{p})^{n-1}}-\frac{1}{(x^{-}_{p})^{n-1}}\right),
\end{equation}
where the Zhukovski variables suitably depend on each particle's representation, we find%
\footnote{%
In fact, if we assume no particle production it is enough to require a single higher charge to be conserved to rule out all but one of  the processes in~\eqref{eq:badprocesses}. Imposing higher conservation laws would force particle number to be conserved, as we have already implicitly assumed.
}
 that the only allowed processes are the ones where $m$ is transmitted along with the momentum:
\begin{equation}
  \label{eq:allowedprocesses}
  (p_1,m_1;\,p_2,m_2)\ \longrightarrow\ (p_1',m_1';\,p_2',m_2')=(p_2,m_2;\,p_1,m_1) .
\end{equation}
As the sign of $m$ determines the left/right flavour, also this label is transmitted. 
This restriction on the scattering processes is compatible with the perturbative calculations so far performed in this theory~\cite{Rughoonauth:2012qd,Sundin:2012gc, Beccaria:2012kb,Sundin:2013ypa,Engelund:2013fja, Abbott:2013ixa,Bianchi:2014rfa,Roiban:2014cia,Sundin:2014ema}. Therefore, for the time being we will work under the assumption that only the processes~\eqref{eq:allowedprocesses} are allowed.

\subsection{Constraining the S~matrix}
Let us consider an arbitrary (super)charge $\gen{Q}$ of $\psu(1|1)^2_{\ce}$ in the two-particle representation, denoted by $\gen{Q}(p_1,p_2)$. This is a $16^2\times16^2$ matrix, which can be decomposed into $2^2\times 2^2$ matrices, corresponding to irreducible two-particle representations of the form $\varrho_1\otimes\varrho_2$, identified by the charges $m_1,m_2$ and by the grading. The possible representations~$\varrho_i$ have been presented in section~\ref{sec:shortreprs}. If we denote such matrices by $\gen{Q}_{m_1,m_2}^{\varrho_1,\varrho_2}(p_1,p_2)$,  we can write down the constraints on the S~matrix
\begin{equation}
  \label{eq:Smatrixinvariance} 
  \gen{Q}_{m_2,m_1}^{\varrho_2,\varrho_1}(p_2,p_1)\,\Smat_{m_1,m_2}^{\varrho_1,\varrho_2}(p_1,p_2)
  =
  \Smat_{m_1,m_2}^{\varrho_1,\varrho_2}(p_1,p_2)\,\gen{Q}_{m_1,m_2}^{\varrho_1,\varrho_2}(p_1,p_2) .
\end{equation}
These equations are similar to the ones solved in reference~\cite{Lloyd:2014bsa} as an auxiliary problem in order to find the $\AdS_3\times\Sphere^3\times\Torus^4$ mixed-flux S~matrix.%
\footnote{%
  Note in fact that the $\psu(1|1)^4_{\ce}$ symmetry of that theory factors precisely into two copies of the $\psu(1|1)^2_{\ce}$ discussed here.}
The situation here is a bit more general, as we want to allow the masses to take real values $0\leq|m|\leq1$. Still it is straightforward to find that each block of the S-matrix is completely determined up to an overall pre-factor---a \emph{dressing factor}. We collect the expressions for these blocks in appendix~\ref{app:Smat}.

There are some further constraints that we should impose for consistency: braiding unitarity imposes
\begin{equation}
  \Smat(p_2,p_1)\;\Smat(p_1,p_2)=\gen{1} ,
\end{equation}
while physical unitarity requires~$\Smat$ to be unitary as a matrix. Both of these constraints will yield restrictions on the dressing factors. More restrictions will follow from requiring crossing invariance~\cite{Janik:2006dc}, as we will describe in section~\ref{sec:constraints-on-dressing-phase}. Finally, for consistency with factorisation of scattering, the Yang-Baxter equation
\begin{equation}
  \Smat(p_2,p_3)\otimes\gen{1}\cdot 
  \gen{1}\otimes\Smat(p_1,p_3)\cdot 
  \Smat(p_1,p_2)\otimes\gen{1}
  =
  \gen{1}\otimes\Smat(p_1,p_2)\cdot 
  \Smat(p_1,p_3)\otimes\gen{1}\cdot 
  \gen{1}\otimes\Smat(p_2,p_3) ,
\end{equation}
must also hold. This is in fact the case for an S~matrix composed of the blocks given in appendix~\ref{app:Smat}.

Another constraint is the discrete \emph{left-right symmetry}~\cite{Borsato:2012ud}. We have seen that the left and right representations of section~\ref{sec:quadraticrepr} (and further detailed in appendix~\ref{app:charges}) are mapped into each other by swapping L$\leftrightarrow$R everywhere. Note than in presence of a non-vanishing NS-NS flux, this also means flipping the sign of~$q$. We will assume that this discrete symmetry still holds at the level of the S~matrix---compatibly with perturbative calculations. This is automatically the case for all blocks we construct, but gives further relations between the dressing factors.

\subsection{Blocks and dressing factors}
Overall, the S~matrix splits into $8\times8=64$ blocks, one for each possible combination of masses and left/right flavours of the incoming particles. In principle, each of those comes with a dressing factor, which cannot be fully determined just by symmetry arguments. However, unitarity and left-right symmetry reduce this number significantly. Let us list such blocks to better investigate how this happens.

\paragraph{Same mass, same chirality.}
Let us consider two particles of mass~$|m|$ and same target-space chiralities. We therefore have eight blocks%
\footnote{%
  To keep the notation manageable we use $\varrho_{\sL}\equiv\text{L}$,  $\varrho_{\sR}\equiv\text{R}$, \textit{etc.} in the S-matrix indices.}
\begin{equation}
  \label{eq:commLL}
  \begin{aligned}
    \Sigma_{m,m}^{\sL\sL}\,\Smat_{m,m}^{\sL\sL},\quad
    \Sigma_{m,m}^{\sR\sR}\,\Smat_{m,m}^{\sR\sR},&\qquad\text{with}\qquad
    |m|=0,\alpha,1-\alpha ,
    \\
    \Sigma_{m,m}^{\sL\sL}\,\Smat_{m,m}^{\tilde{\sL}\tilde{\sL}},\quad
    \Sigma_{m,m}^{\sR\sR}\,\Smat_{m,m}^{\tilde{\sR}\tilde{\sR}},&\qquad\text{with}\qquad
    |m|=1 ,
  \end{aligned}
\end{equation}
where we have multiplied each block by its dressing factor~$\Sigma$. We single out the heavy-mode S~matrix since it scatters representations with a grading that is opposite to the one of light modes. Its matrix structure is related in a simple way to that of the other blocks, see appendix~\ref{app:Smat}.
The matrix part of all blocks depends on the masses only through the Zhukovski parameters $x^\pm_{\sL,\sR}$, and can be found in equations~(\ref{eq:su(1|1)2-Smat-grad1}--\ref{eq:su(1|1)2-Smat-grad2}).
% The heavy-mode S matrices $\Smat_{\mu,\mu}^{\tilde{\sL}\tilde{\sL}},\Smat_{\mu,\mu}^{\tilde{\sR}\tilde{\sR}}$ responsible for the scatetring of heavy modes may be understood as a proper transformation of the previous ones, to take into account the different grading of the representations, see~\eqref{eq:su(1|1)2-Smat-grad2}.

If we assume that the dressing factors are related by LR symmetry, \ie,
\begin{equation}
  \label{eq:commLR}
  \Sigma_{m,m}^{\sL\sL}(p_1,p_2;q)=\Sigma_{m,m}(x_1^{\sL},x_2^{\sL};+q) ,\qquad
  \Sigma_{m,m}^{\sR\sR}(p_1,p_2;q)=\Sigma_{m,m}(x_1^{\sR},x_2^{\sR};-q) ,
\end{equation}
for appropriate functions $\Sigma_{m,m}$, we are then  left with four undetermined factors.

\paragraph{Same mass, opposite chirality.}
In a very similar way we also start out with eight blocks here
\begin{equation}
\begin{aligned}
\Sigma_{m,m}^{\sL\sR}\,\Smat_{m,m}^{\sL\sR},\quad
\Sigma_{m,m}^{\sR\sL}\,\Smat_{m,m}^{\sR\sL},&\qquad\text{with}\qquad
|m|=0,\alpha,1-\alpha ,
\\
\Sigma_{m,m}^{\sL\sR}\,\Smat_{m,m}^{\tilde{\sL}\tilde{\sR}},\quad
\Sigma_{m,m}^{\sR\sL}\,\Smat_{m,m}^{\tilde{\sR}\tilde{\sL}},&\qquad\text{with}\qquad
|m|=1 ,
\end{aligned}
\end{equation}
where the explicit expressions for the two cases are collected in~\eqref{eq:su(1|1)2-Smat-LRgrad1} and~\eqref{eq:su(1|1)2-Smat-LRgrad2} respectively.
Upon imposing LR symmetry we get
\begin{equation}
  \Sigma_{m,m}^{\sL\sR}(p_1,p_2;q)=\widetilde{\Sigma}_{m,m}(x_1^{\sL},x_2^{\sR};+q) , \qquad
  \Sigma_{m,m}^{\sR\sL}(p_1,p_2;q)=\widetilde{\Sigma}_{m,m}(x_1^{\sR},x_2^{\sL};-q) ,
\end{equation}
for four appropriate $\widetilde{\Sigma}_{m,m}$.

\paragraph{Different mass, same chirality.}
We start with 24 blocks. We have to distinguish between the case in which only light modes are involved
\begin{equation}\label{eq:smat-LL-light-diff-mass}
  \begin{aligned}
    \Sigma_{m_1,m_2}^{\sL\sL}\,\Smat_{m_1,m_2}^{\sL\sL},\quad
    \Sigma_{m_1,m_2}^{\sR\sR}\,\Smat_{m_1,m_2}^{\sR\sR},&\qquad\text{with}\qquad
    |m_1|,|m_2|=0,\alpha,1-\alpha\, \quad
    |m_1|\neq|m_2| ,
  \end{aligned}
\end{equation}
and the case in which light modes scatter with heavy ones
\begin{equation}\label{eq:smat-LL-light-heavy}
  \begin{aligned}
    \Sigma_{m_1,m_2}^{\sL\sL}\,\Smat_{m_1,m_2}^{\sL\tilde{\sL}},\quad
    \Sigma_{m_1,m_2}^{\sR\sR}\,\Smat_{m_1,m_2}^{\sR\tilde{\sR}},&\qquad\text{with}\qquad
    |m_1|=0,\alpha,1-\alpha,\quad |m_2|=1,
    \\
    \Sigma_{m_1,m_2}^{\sL\sL}\,\Smat_{m_1,m_2}^{\tilde{\sL}\sL},\quad
    \Sigma_{m_1,m_2}^{\sR\sR}\,\Smat_{m_1,m_2}^{\tilde{\sR}\sR},&\qquad\text{with}\qquad
    |m_1|=1,\quad |m_2|=0,\alpha,1-\alpha.
  \end{aligned}
\end{equation}
In~\eqref{eq:smat-LL-light-diff-mass} we find again S matrices of the form~\eqref{eq:su(1|1)2-Smat-grad1}, since the mass dependence is just encoded in the spectral parameters $x^\pm_{\sL,\sR}$.
The matrices appearing in~\eqref{eq:smat-LL-light-heavy} are instead found in~\eqref{eq:su(1|1)2-Smat-grad3} and~\eqref{eq:su(1|1)2-Smat-grad4}, because of the different grading of the two representations that scatter.

Clearly LR-symmetry halves the amount of independent blocks. In this case it is also interesting to observe that braiding unitarity gives, in the appropriate normalisation of appendix~\ref{app:Smat}
\begin{equation}
\Sigma_{m_2,m_1}^{\sL\sL}(p_2,p_1)\,\Sigma_{m_1,m_2}^{\sL\sL}(p_1,p_2)=1\qquad\text{and}\qquad
\Sigma_{m_2,m_1}^{\sR\sR}(p_2,p_1)\,\Sigma_{m_1,m_2}^{\sR\sR}(p_1,p_2)=1.
\end{equation}
This means that these scalar factors are fixed if we determine the six functions
\begin{equation}
\Sigma_{m_1,m_2}(p_1,p_2;q)\qquad\text{with}\qquad
|m_1|,|m_2|=0,\alpha,1-\alpha,1,\quad
|m_1|<|m_2|.
\end{equation}

\paragraph{Different mass, opposite chirality.}
This case resembles the one above, and we start again with 24 blocks. For light-light scattering we find
\begin{equation}
\Sigma_{m_1,m_2}^{\sL\sR}\,\Smat_{m_1,m_2}^{\sL\sR},\quad
\Sigma_{m_1,m_2}^{\sR\sL}\,\Smat_{m_1,m_2}^{\sR\sL},\qquad\text{with}\qquad
|m_1|,|m_2|=0,\alpha,1-\alpha,\quad
|m_1|\neq|m_2|,
\end{equation}
while light-heavy scattering yields
\begin{equation}
\begin{aligned}
\Sigma_{m_1,m_2}^{\sL\sR}\,\Smat_{m_1,m_2}^{\sL\tilde{\sR}},\quad
\Sigma_{m_1,m_2}^{\sR\sL}\,\Smat_{m_1,m_2}^{\sR\tilde{\sL}},&\qquad\text{with}\qquad
|m|=0,\alpha,1-\alpha, \quad |m|'=1,
\\
\Sigma_{m_1,m_2}^{\sL\sR}\,\Smat_{m_1,m_2}^{\tilde{\sL}\sR},\quad
\Sigma_{m_1,m_2}^{\sR\sL}\,\Smat_{m_1,m_2}^{\tilde{\sR}\sL},&\qquad\text{with}\qquad
|m|=1, \quad|m|'=0,\alpha,1-\alpha.
\end{aligned}
\end{equation}
The relevant S matrices are collected in~\eqref{eq:su(1|1)2-Smat-LRgrad1},~\eqref{eq:su(1|1)2-Smat-LRgrad3} and~\eqref{eq:su(1|1)2-Smat-LRgrad4}.
Here unitarity relates the LR to the RL channel,
\begin{equation}
\Sigma_{m_2,m_1}^{\sL\sR}(p_2,p_1)\,\Sigma_{m_1,m_2}^{\sR\sL}(p_1,p_2)=1,
\end{equation}
and again we are left with six functions
\begin{equation}
\widetilde{\Sigma}_{m_1,m_2}(p_1,p_2;q)\qquad\text{with}\qquad
|m_1|,|m_2|=0,\alpha,1-\alpha,1\,,\quad
|m_1|<|m_2|.
\end{equation}
Let us stress again that all this discussion was done for the case in which we can include the heavy modes in the asymptotic particle spectrum.
Should the heavy modes be composite or be bound states, we would not need to compute their scattering matrices. In that case, we would only have to determine 12~blocks and the relative dressing factors.

\subsection{Constraints on the dressing factors}
\label{sec:constraints-on-dressing-phase}

The matrix part of each block is fixed by requiring equation~\eqref{eq:Smatrixinvariance} to hold for suitably chosen representations $\varrho_1,\,\varrho_2$.
%Once the supercharges have been fixed to be in a representation $\varrho\otimes\varrho'$ where both $\varrho$ and~$\varrho'$ are short representations of definite momentum, mass, chirality and grading, S~matrix is constrained to take the form given in appendix~\ref{app:Smat}.
The normalisation of each block, and hence of the dressing factor, is a matter of convention. Our choices in appendix~\ref{app:Smat} 
%follow the ones of reference~\cite{Borsato:2014hja, Lloyd:2014bsa} and 
aim at simplifying the constraints on the dressing factors. These come from braiding and physical unitarity, and from crossing symmetry.

\paragraph{Constraints from unitarity.}
Braiding unitarity imposes that the dressing factors satisfy
\begin{equation}
\begin{gathered}
\Sigma_{m_2,m_1}^{\sL\sL}(p_2,p_1)\,\Sigma_{m_1,m_2}^{\sL\sL}(p_1,p_2)=1,
\qquad
\Sigma_{m_2,m_1}^{\sR\sR}(p_2,p_1)\,\Sigma_{m_1,m_2}^{\sR\sR}(p_1,p_2)=1,
\\
\Sigma_{m_2,m_1}^{\sL\sR}(p_2,p_1)\,\Sigma_{m_1,m_2}^{\sR\sL}(p_1,p_2)=1,
\end{gathered}
\end{equation}
while physical unitarity yields
\begin{equation}
\begin{aligned}
\left(\Sigma_{m_1,m_2}^{\sL\sL}(p_1,p_2)\right)^{*}\,\Sigma_{m_1,m_2}^{\sL\sL}(p_1,p_2)=1,
\qquad
\left(\Sigma_{m_1,m_2}^{\sR\sR}(p_1,p_2)\right)^{*}\,\Sigma_{m_1,m_2}^{\sR\sR}(p_1,p_2)=1,
\\
\left(\Sigma_{m_1,m_2}^{\sL\sR}(p_1,p_2)\right)^{*}\,\Sigma_{m_1,m_2}^{\sL\sR}(p_1,p_2)=1,
\qquad
\left(\Sigma_{m_1,m_2}^{\sR\sL}(p_1,p_2)\right)^{*}\,\Sigma_{m_1,m_2}^{\sR\sL}(p_1,p_2)=1,
\end{aligned}
\end{equation}
for any choice of the masses $m_1$ and $m_2$, where $*$ denotes complex conjugation. These conditions imply that all the dressing phases are pure phases. 

\paragraph{Constraints from crossing.}
Invariance under the particle-to-antiparticle transformation requires that the S~matrix is compatible with crossing symmetry~\cite{Janik:2006dc}.
On the one-particle representations we might define the charge conjugation matrix as
\begin{equation}
\newcommand{\0}{\color{black!40}0}
  \renewcommand{\arraystretch}{1.1}
  \setlength{\arraycolsep}{3pt}
  \mathscr{C}=\!\left(\!
    \mbox{\footnotesize$
      \begin{array}{cccc}
        \0 & \0 & 1 & \0 \\
        \0 & \0 & \0 & i \\
        1 & \0 & \0 & \0 \\
        \0 & i & \0 & \0 \\
      \end{array}$}\!
  \right) ,
\end{equation}
in the basis%
\footnote{%
The charge conjugation matrix can be chosen to be the same also for the representation $\widetilde{\varrho}_{\sL}\oplus\widetilde{\varrho}_{\sR}$ in the basis $\{\tilde{\phi}^{\sL},\tilde{\psi}^{\sL},\tilde{\phi}^{\sR},\tilde{\psi}^{\sR}\}$.} $\{\phi^{\sL},\psi^{\sL},\phi^{\sR},\psi^{\sR}\}$, where $\phi$ denotes bosons and $\psi$ fermions belonging to a $\algPSU(1|1)^2_{\ce}$ short representation.
After analytically continuing the momentum $p$ to $\bar{p}$ we have to implement crossing on the Zhukovski variables as in reference~\cite{Lloyd:2014bsa}
\begin{equation}
x^\pm_{\sL}(\bar{p})=\frac{1}{x^\pm_{\sR}(p)},
\qquad
x^\pm_{\sR}(\bar{p})=\frac{1}{x^\pm_{\sL}(p)},
\end{equation}
and to resolve square-root ambiguities of equation~\eqref{eq:etadef} by
\begin{equation}
\eta^{\sL}(\bar{p})=\frac{i}{x^+_{\sR}(p)}\eta^{\sR}(p),
\qquad
\eta^{\sR}(\bar{p})=\frac{i}{x^+_{\sL}(p)}\eta^{\sL}(p).
\end{equation}
The crossing equations may be written compactly in terms of the matrix 
\begin{equation}
\mathbf{S}=\Pi\;\mathcal{S},
\end{equation}
where $\Pi$ is the permutation matrix. Then we have, in matrix form,
\begin{equation}\label{eq:cr-matrix-form}
\begin{aligned}
\mathscr{C}_1 \cdot \mathbf{S}^{\text{t}_1}(\bar{p}_1,p_2) \cdot \mathscr{C}_1^{-1}  \cdot \mathbf{S}(p_1,p_2) &= \mathbf{1} ,
\end{aligned}
\end{equation}
where we have used the notation $\mathscr{C}_1 = \mathscr{C}  \otimes \mathbf{1}$, and ${}^{\text{t}_1}$ denotes transposition on the first space.
These equations amount to constraints just on the scalar factors
\begin{equation}\label{eq:explic-cr-eq}
  \begin{aligned}
    \Sigma^{\sR\sL}_{m_1m_2}(x_{\sR}(\bar{p}_1),x_{\sL}(p_2))&\;\Sigma^{\sL\sL}_{m_1m_2}(x_{\sL}(p_1),x_{\sL}(p_2))=c(x_{\sL\, 1},x_{\sL\, 2}) , \\
    \Sigma^{\sL\sL}_{m_1m_2}(x_{\sL}(\bar{p}_1),x_{\sL}(p_2))&\;\Sigma^{\sR\sL}_{m_1m_2}(x_{\sR}(p_1),x_{\sL}(p_2))=\widetilde{c}(x_{\sR\, 1},x_{\sL\, 2}) , \\[8pt]
    \Sigma^{\sL\sR}_{m_1m_2}(x_{\sL}(\bar{p}_1),x_{\sR}(p_2))&\;\Sigma^{\sR\sR}_{m_1m_2}(x_{\sR}(p_1),x_{\sR}(p_2))=c(x_{\sR\, 1},x_{\sR\, 2}) , \\
    \Sigma^{\sR\sR}_{m_1m_2}(x_{\sR}(\bar{p}_1),x_{\sR}(p_2))&\;\Sigma^{\sL\sR}_{m_1m_2}(x_{\sL}(p_1),x_{\sR}(p_2))=\widetilde{c}(x_{\sL\, 1},x_{\sR\, 2}) ,
  \end{aligned}
\end{equation}
where we have defined the functions of the Zhukovski variables
\begin{equation}
  \begin{aligned}
    c(x_1,x_2)=&\left(\frac{x^+_{ 1}}{x^-_{ 1}}\right)^{+1/4}\left(\frac{x^+_{ 2}}{x^-_{ 2}}\right)^{-1/4}\frac{ x^-_{ 1}-x^-_{ 2}}{x^+_{ 1}-x^-_{ 2} }\sqrt{\frac{   x^+_{ 1}-x^+_{ 2}}{   x^-_{ 1}-x^-_{ 2}}},\\
    \widetilde{c}(x_1,x_2)=&\left(\frac{x^+_{ 1}}{x^-_{ 1}}\right)^{-1/4}
    \left(\frac{x^+_{ 2}}{x^-_{ 2}}\right)^{-3/4}
    \frac{
      1-x^-_{ 1}x^+_{ 2}}{1-x^-_{ 1}x^-_{ 2} }
    \sqrt{\frac{1-\frac{1}{x^-_{ 1}x^-_{ 2}}}{1-\frac{1}{x^+_{ 1}x^+_{ 2}}}} .
  \end{aligned}
\end{equation}
Thanks to the normalisations of the S matrices introduced in appendix~\ref{app:Smat}, the crossing equations above take the same form for any choice of the masses $m_1,m_2$.
Furthermore, it is clear that LR symmetry relates the first and third lines, and the second and fourth lines in~\eqref{eq:explic-cr-eq}.%
\footnote{%
Let us stress once more that such normalisations are arbitrary, and that different normalisations would produce different right-hand-sides of the crossing equations.
}

It would be very interesting to solve these crossing equations, at least at the so-called Arutyunov-Frolov-Staudacher (AFS) order of the dressing phases.
For the case of pure R-R ($q=0$), an AFS order of the phases in the massive sector---including scattering of different masses---was recently proposed in~\cite{Abbott:2014pia}.
The proposal of~\cite{Abbott:2014pia} was also shown to be compatible with the crossing equations derived in~\cite{Borsato:2012ud}.
It is easy to see that those crossing equations match with the ones derived here, if we account for the different normalisations on the two sides.
In particular, comparing the (string-frame) S~matrix of~\cite{Borsato:2012ud} with the one constructed here (when we set $q=0$), we see that we have to identify the scalar factors $S^{\sL\sL '},S^{\sR\sL '}$ of~\cite{Borsato:2012ud} with
\begin{equation}
\begin{aligned}
S^{\sL\sL '}(p_1,p_2) &\to \left( \frac{x^+_1}{x^-_1} \right)^{-1/2}\left( \frac{x^+_2}{x^-_2} \right)^{1/2}\ \Sigma^{\sL\sL}_{m_1m_2}(p_1,p_2)\,,\\
S^{\sR\sL '}(p_1,p_2) &\to \left( \frac{x^+_1}{x^-_1} \right)^{-1/4}\left( \frac{x^+_2}{x^-_2} \right)^{1/4}\ \Sigma^{\sR\sL}_{m_1m_2}(p_1,p_2)\,,
\end{aligned}
\end{equation}
where the labels L and R on the Zhukovski variables can be omitted, because $q=0$.
With this identification  we can check\footnote{The equations of~\cite{Borsato:2012ud} are written for crossing in the second variable. For this reason we also need to use braiding unitarity to rewrite them when the first variable is crossed.} that (5.46) of~\cite{Borsato:2012ud} is compatible with~\eqref{eq:explic-cr-eq}.

To conclude, let us comment on the form of the charge-conjugation matrix~\eqref{eq:cr-matrix-form} with respect to the one of references~\cite{Borsato:2014hja,Lloyd:2014bsa}. There, charge conjugation for massless particles involved momentum-dependent expressions of the form~$\text{sgn}(\text{sin}\tfrac{p}{2})$. This is simply because, when taking the $\alpha\to1$ limit and identifying the massless modes here with the ones of~$\AdS_3\times\Sphere^3\times\Torus^4$, a momentum-dependent change of basis is necessary to make the $\so(4)$ symmetry of that case manifest. This is precisely the change of basis discussed  in section~\ref{sec:massless-reps}.

\section{Conclusions}

The all-loop worldsheet S~matrix of the maximally supersymmetric string theory $\AdS_3\times \mathcal{M}_7$ backgrounds can be fixed, up to dressing phases, by determining the off-shell symmetry algebra $\mathcal{A}$ and its representations. This approach, originally used in the context of $\AdS_5\times \Sphere^5$~\cite{Arutyunov:2006ak}, has been particularly useful in the context of type IIB strings on $\AdS_3\times\Sphere^3\times \Torus^4$~\cite{Borsato:2014exa,Borsato:2014hja,Lloyd:2014bsa}.  Unlike the  $\AdS_5\times \Sphere^5$ background where the two formulations are equivalent, in the case of string theory on $\AdS_3$ it is necessary to use the GS action, rather than the coset action~\cite{Babichenko:2009dk}. This is because the massless fermions that appear in $\AdS_3$ backgrounds do not have conventional kinetic terms in the coset formulation~\cite{Borsato:2014exa}. While this makes the computations more involved than in the case of $\AdS_5$, a major advantage of this approach is that it treats massive and massless modes democratically, circumventing previous problems associated with incorporating massless modes into the integrability construction. 

In this paper we have applied this method to type IIB string theory on the background $\AdS_3\times\Sphere^3\times\Sphere^3\times\Sphere^1$ with mixed NS-NS and R-R flux by computing the all-loop worldsheet S matrix between all possible one-particle representations.
The S matrix was fixed, up to dressing factors, using $\mathcal{A}$ and its representations. As we have discussed, it may well be that the heavy modes should not be treated as fundamental particles, and that therefore the relative S matrix needs not to be computed in this way. While we are left with several dressing factors, most of those are related to each other by unitarity and symmetry under left and right or $\alpha\leftrightarrow(1-\alpha)$ exchange. When we consider the scattering of light massive and massless modes, we are left with nine factors: four correspond to light--light massive scattering of same or different mass and same or opposite chirality, two to the scattering of a light massive mode with one of the two massless representations, and three to massless scattering of each of the two representations with itself and with each other. Should we treat the heavy modes as fundamental, we would have to consider six  additional dressing factors. We have
also determined the crossing relations that all dressing factors should satisfy.

Because we have been working with the GS action, the methods used here are quite robust and do not, for example, depend on the background being a semi-symmetric space. As a result, they could be applied to other, less symmetric backgrounds, associated not just to less-symmetric cosets~\cite{Wulff:2014kja,Hoare:2014kma}, but perhaps also to other $\AdS$ backgrounds such as~\cite{Maldacena:2000mw,Gaiotto:2009gz,ReidEdwards:2010qs,Aharony:2012tz,Apruzzi:2014qva}.
While these latter backgrounds are not expected to be integrable, it would be interesting to establish what happens to the symmetry algebra of the gauge-fixed action when the 
level-matching condition is relaxed. In the case of the integrable backgrounds studied in this paper and previous works, the Lie-algebra structure is preserved and the algebra is merely centrally extended to $\algA$ when one goes off-shell. One may wonder whether this relatively simple structure is a result of the underlying integrability of the theory and whether it will be significantly modified in more generic backgrounds. 

The S~matrix we have constructed gives rise to a three-parameter family of quantum integrable models, controlled by $\lambda$, $\alpha$ and $q$. Together with the presence of massless modes this provides a rich setting for investigating more fully the integrability structures present here. It would be very interesting to understand, for example, the asymptotic Bethe ansatz~\cite{Beisert:2005fw}, finite-gap equations~\cite{Babichenko:2009dk,OhlssonSax:2011ms,Lloyd:2013wza,Babichenko:2014yaa},
the thermodynamic Bethe ansatz~\cite{Ambjorn:2005wa,Arutyunov:2007tc,Arutyunov:2009zu,Gromov:2009tv,Bombardelli:2009ns,Arutyunov:2009ur,Cavaglia:2010nm}, Yangian symmetries~\cite{Beisert:2007ds,Spill:2008tp,Beisert:2014hya,Pittelli:2014ria,Regelskis:2015xxa}, and quantum spectral curve~\cite{Gromov:2013pga,Cavaglia:2014exa,Gromov:2014caa} for these models. The study of boundary integrable boundary conditions for these models is another interesting direction, which has recently been investigated in~\cite{Prinsloo:2015apa}.

The $\AdS$ side of the $\AdS_3/\CFT_2$ correspondence is now likely to be understandable using integrable holography methods. Recently, some signs of integrability on the $\CFT$ side have also been identified~\cite{Sax:2014mea} in the $\CFT$ dual to strings on $\AdS_3\times\Sphere^3\times \Torus^4$. Much less is known about the $\CFT_2$ dual of strings on $\AdS_3\times\Sphere^3\times\Sphere^3\times\Sphere^1$~\cite{Gauntlett:1998kc,Boonstra:1998yu,Gukov:2004ym,Tong:2014yna}. It would be interesting to see if the $\AdS$ integrability results already known for this background can shed some light on the dual $\CFT_2$. The $\AdS_3/\CFT_2$ correspondence has been recently investigated in the higher-spin limit (see ~\cite{Gaberdiel:2013vva,Gaberdiel:2014cha,Baggio:2015jxa} and references therein). It was found that in this context the $\alpha\rightarrow 0$ limit provided valuable information about the theories, much as it had done in~\cite{OhlssonSax:2011ms,Sax:2012jv}. Since it is widely expected that the higher-spin theory should arise as a tensionless limit of the string backgrounds it would be very interesting to see the precise way in which these can be related. Perhaps the large symmetries (Yangian and W-algebra, respectively) can be used in this context.

\section*{Acknowledgements}
We would like to thank Gleb Arutyunov, Sergey Frolov, Ben Hoare, Tom Lloyd, Alessandro Torrielli, Arkady Tseytlin,  Linus Wulff and Kostya Zarembo for helpful discussions.
R.B.\@ acknowledges support by the Netherlands Organization for Scientific Research (NWO) under the VICI grant 680-47-602. 
His work is also part of the ERC Advanced grant research programme No.\ 246974,
 ``Supersymmetry: a window to non-perturbative physics'',
and of the D-ITP consortium, a program of the NWO that is funded by the Dutch Ministry of Education, Culture and Science (OCW).
O.O.S.'s  work was supported by the ERC Advanced grant No.~290456,
 ``Gauge theory -- string theory duality''.
A.S.'s work is funded by the People Programme (Marie Curie Actions) of the European Union, Grant Agreement No.~317089 (GATIS). A.S. also acknowledges the hospitality at APCTP where part of this work was
done.
B.S.\@ acknowledges funding support from an STFC Consolidated Grant  ``Theoretical Physics at City University''
ST/J00037X/1. BS would also like to thank Matthias Staudacher and Humboldt University for hospitality during the final stages of this project.

%\newpage
%
\appendix

\section{Conventions}
\label{sec:conventions}

For $\AdS_3$ and $\Sphere^3$ we consider the three-dimensional gamma matrices\footnote{%
  Our conventions are the same as those of~\cite{Babichenko:2009dk}, except for the definition of $\gamma^0$ and $\gamma^2$.
}%
\begin{equation}
  \begin{aligned}
    \gamma^0 &= -i\sigma_3 , \quad &
    \gamma^1 &= \sigma_1 , \quad &
    \gamma^2 &= \sigma_2 ,\\
    \gamma^3 &= \sigma_1 , \quad &
    \gamma^4 &= \sigma_2 , \quad &
    \gamma^5 &= \sigma_3 , \\
    \gamma^6 &= \sigma_1 , \quad &
    \gamma^7 &= \sigma_2 , \quad &
    \gamma^8 &= \sigma_3 .
  \end{aligned}
\end{equation}
The ten-dimensional gamma matrices are then given by
\begin{equation}
  \newcommand{\J}{\mathrlap{\,\mathds{1}}\hphantom{\sigma_a}}
  \newcommand{\JJ}{\mathrlap{\,\mathds{1}}\hphantom{\gamma^A}}
  \begin{aligned}
    \Gamma^A &= +\sigma_1 \otimes \sigma_2 \otimes \gamma^A \otimes \JJ \otimes \JJ , & A&= 0,1,2, \\
    \Gamma^A &= +\sigma_1 \otimes \sigma_1 \otimes \JJ \otimes \gamma^A \otimes \JJ , & A&= 3,4,5, \\
    \Gamma^A &= +\sigma_1 \otimes \sigma_3 \otimes \JJ \otimes \JJ \otimes \gamma^A , & A&= 6,7,8, \\
    \Gamma^9 &= -\sigma_2 \otimes \J \otimes \JJ \otimes \JJ \otimes \JJ .
  \end{aligned}
\end{equation}
We then have
\begin{equation}
  \newcommand{\J}{\mathrlap{\,\mathds{1}}\hphantom{\sigma_a}}
  \begin{aligned}
    \Gamma^{05} &= \phantom{i} {-}\J \otimes \sigma_3  \otimes \sigma_3 \otimes \sigma_3 \otimes \J , \\
    \Gamma^{012} &= \phantom{i} {+} \sigma_1 \otimes \sigma_2  \otimes \J \otimes \J \otimes \J , \\
    \Gamma^{345} &= +i \sigma_1 \otimes \sigma_1  \otimes \J \otimes \J \otimes \J , \\
%    \Gamma^{678} &= +i \sigma_1 \otimes \sigma_3  \otimes \I \otimes \I \otimes \I , \\
    \Gamma^{012345} &= +\phantom{i} \J \otimes \sigma_3  \otimes \J \otimes \J \otimes \J , \\
    \Gamma^{1234} &= -\phantom{i} \J \otimes  \J \otimes  \sigma_3 \otimes  \sigma_3 \otimes  \J , \\
    \Gamma^{6789} &= +\phantom{i} \sigma_3 \otimes \sigma_3 \otimes \J \otimes \J \otimes \J , \\
    \Gamma = \Gamma^{0123456789} &= \phantom{i} {+} \sigma_3 \otimes \J \otimes \J \otimes \J \otimes \J .
  \end{aligned}
\end{equation}
The gamma matrices satisfy
\begin{equation}
  (\Gamma^A)^t = - T \Gamma^A T^{-1} , \qquad
  (\Gamma^A)^\dag = - C \Gamma^A C^{-1} , \qquad
  (\Gamma^A)^* = + B \Gamma^A B^{-1} , \qquad
\end{equation}
where
\begin{equation}
  T = -i\sigma_2 \otimes \sigma_2 \otimes \sigma_2 \otimes \sigma_2 \otimes \sigma_2 , \qquad
  C = \Gamma^0 , \qquad
  B = -\Gamma^0 \, T ,
\end{equation}
Note that
\begin{equation}
  \begin{gathered}
    T^\dag T = C^\dag C = B^\dag B = 1 , \qquad
    B^t = T C^\dag , \\
    T^\dag = - T = + T^t , \qquad
    C^\dag = - C = + C^t , \qquad
    B^\dag = + B = + B^t , \\
    T = - \Gamma^{01479} , \qquad
    C = -i \sigma_1 \otimes \sigma_2 \otimes \sigma_3 \otimes \mathds{1} \otimes \mathds{1} , \\
    B = + \sigma_3 \otimes \mathds{1} \otimes \sigma_1 \otimes \sigma_2 \otimes \sigma_2 = - \Gamma^{1479} , \\
    B \Gamma B^\dag = \Gamma^* .
  \end{gathered}
\end{equation}
The Majorana spinors satisfy the conditions
\begin{equation}
  \theta^* = B \theta , \qquad
  \bar{\theta} = \theta^\dag C = \theta^t T .
\end{equation}

\section{Killing spinors}
\label{sec:Killing-spinors}

The background is supported by Ramond-Ramond three-form flux satisfying
\begin{equation}
  \slashed{F} = 12 ( \Gamma^{012} + \cos\varphi \, \Gamma^{345} + \sin\varphi \, \Gamma^{678} ) .
\end{equation}
Let us introduce the matrices\footnote{%
  The gamma matrices that appear in this equations are the ones that were defined in section~\ref{sec:conventions}. Note that the summations over $i$ runs over different values in the various terms, corresponding to the coordinates $z_1$, $z_2$, $\tilde{y}_3$, $\tilde{y}_4$, $\tilde{x}_6$ and $\tilde{x}_7$.
}%
\begin{equation}
  \begin{aligned}
    \hat{M} &= 
    \frac{ 1 - \tfrac{1}{2} z_i \Gamma^{012} \Gamma^i }{ \sqrt{1 - \frac{z^2}{4}} }
    \frac{ 1 - \tfrac{1}{2} \tilde{y}_i \Gamma^{345} \Gamma^i }{ \sqrt{1 + \frac{\tilde{y}^2}{4}} }
    \frac{ 1 - \tfrac{1}{2} \tilde{x}_i \Gamma^{678} \Gamma^i }{ \sqrt{1 + \frac{\tilde{x}^2}{4}} }
    e^{ - \frac{t}{2} \Gamma^{12} - \frac{\tilde{\phi}_5}{2} \Gamma^{34} - \frac{\tilde{\phi}_8}{2} \Gamma^{67} } ,
    \\
    \check{M} &= 
    \frac{ 1 + \tfrac{1}{2} z_i \Gamma^{012} \Gamma^i }{ \sqrt{1 - \frac{z^2}{4}} }
    \frac{ 1 + \tfrac{1}{2} \tilde{y}_i \Gamma^{345} \Gamma^i }{ \sqrt{1 + \frac{\tilde{y}^2}{4}} }
    \frac{ 1 + \tfrac{1}{2} \tilde{x}_i \Gamma^{678} \Gamma^i }{ \sqrt{1 + \frac{\tilde{x}^2}{4}} }
    e^{ + \frac{t}{2} \Gamma^{12} + \frac{\tilde{\phi}_5}{2} \Gamma^{34} + \frac{\tilde{\phi}_8}{2} \Gamma^{67} } ,
  \end{aligned}
\end{equation}
and further define the matrices $M_0$ and $M_t$ as
\begin{equation}
  \label{eq:M0-Mt-definition}
  \begin{aligned}
    \hat{M}(z_i,y_i,x_i,t,\phi_5,\phi_8) &\equiv M_0(z_i,y_i,x_i)M_t(t,\phi_5,\phi_8),
    \\
    \check{M}(z_i,y_i,x_i,t,\phi_5,\phi_8) &\equiv M^{-1}_0(z_i,y_i,x_i)M^{-1}_t(t,\phi_5,\phi_8).
  \end{aligned}
\end{equation}
In the above, for compactness, we have used the rescaled coordinates
\begin{equation}
    \tilde{y}_i = \cos\varphi \, y_i , \qquad
    \tilde{\phi}_5 = \cos\varphi \, \phi_5 , \qquad
    \tilde{x}_i = \sin\varphi \, x_i , \qquad
    \tilde{\phi}_8 = \sin\varphi \, \phi_8 .
\end{equation}
We further introduce the rotated vielbeins
\begin{equation}
  \slashed{\hat{E}}_m = \hat{M}^{-1} \slashed{E}_m \hat{M} ,
  \qquad
  \slashed{\check{E}}_m = \check{M}^{-1} \slashed{E}_m \check{M} ,
\end{equation}
and the orthogonal projectors $\Pi_{\pm}$
\begin{equation}
  \Pi_{\pm} = \frac{1}{2} ( 1 \pm \cos\varphi \, \Gamma^{012345} \pm \sin\varphi \, \Gamma^{012678} ) .
\end{equation}
The covariant derivative in the rotated frame can then be written as
\begin{equation}
  \begin{aligned}
    \partial_m + \tfrac{1}{4} \slashed{\wh}_m + \tfrac{1}{48} \slashed{F} \slashed{\hat{E}}_m
    = \partial_m - &\tfrac{1}{2} \slashed{\hat{E}}_m \Gamma^{012}  \Pi_{+} ,
    \\
    \partial_m + \tfrac{1}{4} \slashed{\wc}_m - \tfrac{1}{48} \slashed{F} \slashed{\check{E}}_m
    = \partial_m + &\tfrac{1}{2} \slashed{\check{E}}_m \Gamma^{012}  \Pi_{+} .
  \end{aligned}
\end{equation}
The Killing spinor equations
\begin{equation}
  \bigl( \partial_m + \tfrac{1}{4} \slashed{\omega}_m + \tfrac{1}{48} \slashed{F} \slashed{E}_m \bigr) \epsilon_1 = 0 , \qquad
  \bigl( \partial_m + \tfrac{1}{4} \slashed{\omega}_m - \tfrac{1}{48} \slashed{F} \slashed{E}_m \bigr) \epsilon_2 = 0 ,
\end{equation}
hence have the solutions
\begin{equation}
  \epsilon_1 = \Pi_{-} \hat{M} \epsilon_1^{(0)} ,
  \qquad
  \epsilon_2 = \Pi_{-} \check{M} \epsilon_2^{(0)} ,
\end{equation}
where $\epsilon_i^{(0)}$ are constant spinors.

\section{Components of the spinors \texorpdfstring{$\theta_I$}{theta}}
\label{app:spinors-and-grassmanns}
The Lagrangian and supercurrents presented in sections~\ref{sec:GS-action-ferm-coords} and~\ref{sec:supercurrents} are written in terms of (eight complex) fermionic components $\theta_{Ii}$. The 32-component Majorana-Weyl spinors $\theta_I$ are given in terms of the components of the spinors $\theta_I$ by
\begin{equation}
  \begin{aligned}
    \theta_1 &=
    \frac{1}{2}
    \begin{pmatrix}
      + e^{-i\pi/4} \sin\!\frac{\varphi}{2} \,\theta_{14} \\ 
      + e^{-i\pi/4} \sin\!\frac{\varphi}{2} \,\theta_{13} \\ 
      - e^{-i\pi/4} \cos\!\frac{\varphi}{2} \,\theta_{12} \\ 
      + e^{+i\pi/4} \cos\!\frac{\varphi}{2} \,\theta_{11} \\
      - e^{-i\pi/4} \cos\!\frac{\varphi}{2} \,\bar{\theta}_{11} \\ 
      - e^{+i\pi/4} \cos\!\frac{\varphi}{2} \,\bar{\theta}_{12} \\ 
      + e^{+i\pi/4} \sin\!\frac{\varphi}{2} \,\bar{\theta}_{13} \\ 
      - e^{+i\pi/4} \sin\!\frac{\varphi}{2} \,\bar{\theta}_{14}
    \end{pmatrix}
    \oplus
    \begin{pmatrix}
      + e^{-i\pi/4} \cos\!\frac{\varphi}{2} \,\theta_{14} \\ 
      - e^{-i\pi/4} \cos\!\frac{\varphi}{2} \,\theta_{13} \\ 
      - e^{-i\pi/4} \sin\!\frac{\varphi}{2} \,\theta_{12} \\ 
      - e^{+i\pi/4} \sin\!\frac{\varphi}{2} \,\theta_{11} \\
      + e^{-i\pi/4} \sin\!\frac{\varphi}{2} \,\bar{\theta}_{11} \\ 
      - e^{+i\pi/4} \sin\!\frac{\varphi}{2} \,\bar{\theta}_{12} \\ 
      - e^{+i\pi/4} \cos\!\frac{\varphi}{2} \,\bar{\theta}_{13} \\ 
      - e^{+i\pi/4} \cos\!\frac{\varphi}{2} \,\bar{\theta}_{14}
    \end{pmatrix}
    \oplus
    \begin{pmatrix}
      0 \\ 0 \\ 0 \\ 0 \\ 0 \\ 0 \\ 0 \\ 0
    \end{pmatrix}
    \oplus
    \begin{pmatrix}
      0 \\ 0 \\ 0 \\ 0 \\ 0 \\ 0 \\ 0 \\ 0
    \end{pmatrix} ,
    \\
    \theta_2 &=
    \frac{1}{2}
    \begin{pmatrix}
      - e^{+i\pi/4} \sin\!\frac{\varphi}{2} \,\theta_{24} \\ 
      - e^{+i\pi/4} \sin\!\frac{\varphi}{2} \,\theta_{23} \\ 
      + e^{+i\pi/4} \cos\!\frac{\varphi}{2} \,\theta_{22} \\ 
      + e^{-i\pi/4} \cos\!\frac{\varphi}{2} \,\theta_{21} \\
      - e^{+i\pi/4} \cos\!\frac{\varphi}{2} \,\bar{\theta}_{21} \\ 
      + e^{-i\pi/4} \cos\!\frac{\varphi}{2} \,\bar{\theta}_{22} \\ 
      - e^{-i\pi/4} \sin\!\frac{\varphi}{2} \,\bar{\theta}_{23} \\ 
      + e^{-i\pi/4} \sin\!\frac{\varphi}{2} \,\bar{\theta}_{24}
    \end{pmatrix}
    \oplus
    \begin{pmatrix}
      - e^{+i\pi/4} \cos\!\frac{\varphi}{2} \,\theta_{24} \\ 
      + e^{+i\pi/4} \cos\!\frac{\varphi}{2} \,\theta_{23} \\ 
      + e^{+i\pi/4} \sin\!\frac{\varphi}{2} \,\theta_{22} \\ 
      - e^{-i\pi/4} \sin\!\frac{\varphi}{2} \,\theta_{21} \\
      + e^{+i\pi/4} \sin\!\frac{\varphi}{2} \,\bar{\theta}_{21} \\ 
      + e^{-i\pi/4} \sin\!\frac{\varphi}{2} \,\bar{\theta}_{22} \\ 
      + e^{-i\pi/4} \cos\!\frac{\varphi}{2} \,\bar{\theta}_{23} \\ 
      + e^{-i\pi/4} \cos\!\frac{\varphi}{2} \,\bar{\theta}_{24}
    \end{pmatrix}
    \oplus
    \begin{pmatrix}
      0 \\ 0 \\ 0 \\ 0 \\ 0 \\ 0 \\ 0 \\ 0
    \end{pmatrix}
    \oplus
    \begin{pmatrix}
      0 \\ 0 \\ 0 \\ 0 \\ 0 \\ 0 \\ 0 \\ 0
    \end{pmatrix} ,
  \end{aligned}
\end{equation}
where $\bar{\theta}_{Ii}$ is the complex conjugate of $\theta_{Ii}$.

\section{Cubic order terms}
\label{app:cubic}

In this appendix we collect the cubic order corrections to the fermionic Lagrangian and the supercurrents. Note that these terms all vanish for $\varphi = 0$ and $\varphi = \pi/2$, which is expected since the gauge-fixed Green-Schwarz action for $\AdS_3 \times \Sphere^3 \times \Torus^4$ contains no cubic terms~\cite{Borsato:2014exa,Borsato:2014hja,Lloyd:2014bsa}. The cubic part of the Lagrangian is given by
\begin{equation}
  \begin{aligned}
    \mathcal{L}_F\bigr|_{\text{cubic}} =
    - \sin\varphi \, \cos\varphi \Big(
    & (\bar{\theta}_{12} \theta_{12} - \bar{\theta}_{13} \theta_{13}) ( \dot{\psi} - q \pri{\psi})
    + ( \bar{\theta}_{22} \theta_{22} - \bar{\theta}_{23} \theta_{23}) ( \dot{\psi} + q \pri{\psi} ) 
    \\
    - \tilde{q} & ( 
    \bar{\theta}_{11} \pri{\theta}_{22} + \bar{\theta}_{21} \pri{\theta}_{12} + i \bar{\theta}_{13} \pri{\theta}_{24} + i \bar{\theta}_{23} \pri{\theta}_{14}
    ) X
    \\
    + \tilde{q} & ( 
    \bar{\theta}_{12} \pri{\theta}_{21} + \bar{\theta}_{22} \pri{\theta}_{11} - i \bar{\theta}_{14} \pri{\theta}_{23} - i \bar{\theta}_{24} \pri{\theta}_{13}
    ) \bar{X}
    \\
    + \tilde{q} & ( 
    \bar{\theta}_{11} \pri{\theta}_{23} + \bar{\theta}_{21} \pri{\theta}_{13} - i \bar{\theta}_{12} \pri{\theta}_{24} - i \bar{\theta}_{21} \pri{\theta}_{14}
    ) Y
    \\
    - \tilde{q} & ( 
    \bar{\theta}_{13} \pri{\theta}_{21} + \bar{\theta}_{23} \pri{\theta}_{11} + i \bar{\theta}_{14} \pri{\theta}_{22} + i \bar{\theta}_{24} \pri{\theta}_{12}
    ) \bar{Y}
    \\
    - \tilde{q} & ( \theta_{12} \theta_{23} + \theta_{13} \theta_{22} ) \pri{Z}
    + \tilde{q} ( \bar{\theta}_{12} \bar{\theta}_{23} + \bar{\theta}_{13} \bar{\theta}_{22} ) \pri{\bar{Z}}
    \\
    - \tilde{q} & ( \theta_{11} \theta_{22} - \theta_{12} \theta_{21} ) \pri{Y}
    + i \tilde{q} ( \bar{\theta}_{13} \bar{\theta}_{24} - \bar{\theta}_{14} \bar{\theta}_{23} ) \pri{\bar{Y}}
    \\
    - \tilde{q} & ( \theta_{11} \theta_{23} - \theta_{13} \theta_{21} ) \pri{X}
    + i \tilde{q} ( \bar{\theta}_{12} \bar{\theta}_{24} - \bar{\theta}_{14} \bar{\theta}_{22} ) \pri{\bar{X}}
    \\
    + \tilde{q} & (
    \bar{\theta}_{12} \theta_{22} - \bar{\theta}_{13} \theta_{23} 
    + \bar{\theta}_{22} \theta_{12} - \bar{\theta}_{23} \theta_{13} 
    ) \pri{w}
    \bigr) .
  \end{aligned}
\end{equation}
Note that for $q=1$ only the first line of this expression remains. Furthermore, in that case $\theta_{1i}$ couples to the right-moving part of $\psi$, and $\theta_{2i}$ couples to the right-moving part.

The cubic order corrections to the components of the current $j_{\sL}$ is given by
\begin{equation} % Corresponds to J3 in the Mathematica file
  \begin{aligned}
    j_{\sL}^{\tau} \bigr|_{\text{cubic}} = \tfrac{1}{2}\sin\varphi\cos\varphi e^{-i\pi/4} e^{+ix^-} \smash{\bigl(} &
    + i \bar{\theta}_{11} ( P_Y Y - P_{\bar{Y}} \bar{Y} - P_X X + P_{\bar{X}} \bar{X} )
    \\ &
    - ( \bar{\theta}_{12} \bar{X} - \bar{\theta}_{13} \bar{Y} ) P_{\psi}
    + i\tilde{q} \bar{\theta}_{24} ( \bar{Y} \pri{\bar{X}} + \pri{\bar{Y}} \bar{X} )
    \\ &
    - \tilde{q} ( \bar{\theta}_{22} \bar{X} - \bar{\theta}_{23} \bar{Y} ) \pri{w}
    - \tilde{q} ( \theta_{22} \bar{Y} + \theta_{23} \bar{X} ) \bar{Z}
    \\ &
    - \frac{\tilde{q}}{2} \bar{\theta}_{21} ( \bar{Y}\pri{Y} + \pri{\bar{Y}}Y - \bar{X}\pri{X} -\pri{\bar{X}}X )
    \smash{\bigr)}
  \end{aligned}
\end{equation}
and
\begin{equation}
  \begin{aligned}
    j_{\sL}^{\sigma} \bigr|_{\text{cubic}} = \tfrac{1}{2} e^{-i\pi/4} \sin\varphi\cos\varphi e^{+ix^-} \smash{\bigl(}\!\! &
    -i \tilde{q} (\theta_{22} \bar{Y} + \theta_{23} \bar{X}) ( 2iP_{\bar{Z}} - Z )
    \\ &
    - \tilde{q} \theta_{24} ( 2iP_Y \bar{X} + 2iP_X \bar{Y} + \bar{X}\bar{Y} )
    \\ &
    - q ( \bar{\theta}_{12} \bar{X} - \bar{\theta}_{13} \bar{Y} ) P_{\psi}
    + \tilde{q} ( \bar{\theta}_{22} \bar{X} - \bar{\theta}_{23} \bar{Y} ) P_w
    \\ &
    + ( \tilde{q} \bar{\theta}_{21} + iq \bar{\theta}_{11} ) ( P_Y Y - P_X X )
    \\ &
    + ( \tilde{q} \bar{\theta}_{21} - iq \bar{\theta}_{11} ) ( P_{\bar{Y}} \bar{Y} - P_{\bar{X}} \bar{X} )
    \\ &
    - i\tilde{q} \bar{\theta}_{21} ( \cos^2\varphi \, \bar{Y}Y - \sin^2\varphi \, \bar{X}X )
    \\ &
    + \tilde{q} ( \tilde{q} \bar{\theta}_{12} - iq \bar{\theta}_{22} ) \bar{X} \pri{\psi} 
    - \tilde{q} ( \tilde{q} \bar{\theta}_{13} - iq \bar{\theta}_{23} ) \bar{Y} \pri{\psi} 
    \\ &
    + \frac{i\tilde{q}}{2} ( \tilde{q} \bar{\theta}_{21} - iq \bar{\theta}_{11} ) ( \bar{Y} \pri{Y} - \pri{\bar{Y}} Y - \bar{X} \pri{X} - \pri{\bar{X}} X )
    \smash{\bigr)} .
  \end{aligned}
\end{equation}
Similarly, the corrections to $j_{\sR}$ are given by
\begin{equation} % Corresponds to J2 in the Mathematica file
  \begin{aligned}
    j_{\sR}^{\tau} \bigr|_{\text{cubic}} = \tfrac{1}{2}\sin\varphi\cos\varphi e^{-i\pi/4} e^{+ix^-} \smash{\bigl(} &
    + i \theta_{21} ( P_Y Y - P_{\bar{Y}} \bar{Y} - P_X X + P_{\bar{X}} \bar{X} )
    \\ &
    - ( \theta_{22} X - \theta_{23} Y ) P_{\psi}
    + i\tilde{q} \theta_{14} ( Y \pri{X} + \pri{Y} X )
    \\ &
    + \tilde{q} ( \theta_{12} X - \theta_{13} Y ) \pri{w}
    - \tilde{q} ( \bar{\theta}_{12} Y + \bar{\theta}_{13} X ) \pri{\bar{Z}}
    \\ &
    - \frac{\tilde{q}}{2} \theta_{11} ( \bar{Y}\pri{Y} + \pri{\bar{Y}}Y - \bar{X}\pri{X} -\pri{\bar{X}}X )
    \smash{\bigr)} ,
  \end{aligned}
\end{equation}
and
\begin{equation}
  \begin{aligned}
    j_{\sR}^{\sigma} \bigr|_{\text{cubic}} = \tfrac{1}{2} e^{-i\pi/4} \sin\varphi\cos\varphi e^{+ix^-} \smash{\bigl(}\!\! &
    -i \tilde{q} (\bar{\theta}_{12} Y + \bar{\theta}_{13} X) ( 2iP_Z - \bar{Z} )
    \\ &
    - \tilde{q} \bar{\theta}_{14} ( 2iP_{\bar{Y}} X + 2iP_{\bar{X}} Y + XY )
    \\ &
    + q ( \theta_{22} X - \theta_{23} Y ) P_{\psi}
    - \tilde{q} ( \theta_{12} X - \theta_{13} Y ) P_w
    \\ &
    - ( \tilde{q} \theta_{11} + iq \theta_{21} ) ( P_Y Y - P_X X )
    \\ &
    - ( \tilde{q} \theta_{11} - iq \theta_{21} ) ( P_{\bar{Y}} \bar{Y} - P_{\bar{X}} \bar{X} )
    \\ &
    + i\tilde{q} \theta_{11} ( \cos^2\varphi \, \bar{Y}Y - \sin^2\varphi \, \bar{X}X )
    \\ &
    + \tilde{q} ( \tilde{q} \theta_{22} + iq \theta_{12} ) X \pri{\psi} 
    - \tilde{q} ( \tilde{q} \theta_{23} + iq \theta_{13} ) Y \pri{\psi} 
    \\ &
    + \frac{i\tilde{q}}{2} ( \tilde{q} \theta_{21} + iq \theta_{11} ) ( \bar{Y} \pri{Y} - \pri{\bar{Y}} Y - \bar{X} \pri{X} - \pri{\bar{X}} X )
    \smash{\bigr)} .
  \end{aligned}
\end{equation}

\section{Quadratic charges}
\label{app:charges}
In this appendix we will spell out the supercharges at quadratic order in the fields, and fix the conventions to cast them in the oscillator notation~\eqref{eq:supercharges}.

\subsection{Expression in terms of fields}
Let us introduce complex combinations of the fields. For the bosons we choose

\begin{center}
  \begin{tabular}{cccc}
    \toprule
    $|m|=1$ & $|m|=\alpha$ & $|m|=1-\alpha$ & $|m|=0$ \\
    \midrule
    $Z = -z_2 + i z_1$ & $Y = -y_3 - i y_4$ & $X= -x_6 -i x_7$ & $W = w - i \psi$ \\
    $\bar{Z} = -z_2 - i z_1$ & $\bar{Y} = -y_3 + i y_4$ & $\bar{X} = -x_6 +i x_7$ & $\bar{W} = w + i \psi$ \\
    \bottomrule
  \end{tabular}
\end{center}
Note that at quadratic order we can combine the massless coordinate
$w$ coming from $\Sphere^1$ with $\psi$ coming from the combination of the equators of the two three-spheres. The conjugate momenta $P_Z, P_{\bar{Z}},$ \textit{etc.}, are define in such a way to have canonical Poisson brackets with the fields, so that $P_{Z}=\tfrac{1}{2}\partial_0\bar{Z}$, $P_{\bar{Z}}=\tfrac{1}{2}\partial_0Z$ and so on. In this way, we have the canonical commutation relations\footnote{Our convention for the bars on $P_{{Z}},P_{\bar{Z}}$, \textit{etc.} is different from the one of~\cite{Borsato:2014hja, Lloyd:2014bsa}.}
\begin{equation}
  \big[
  Z(x),P_{Z}(y)
  \big]=i\,\delta(x-y),
\end{equation}
and so on.

We also redefine the fermions, denoting them by $\theta^{\sL\, j}, \theta^{\sR\, j}$ and indicating their complex conjugates by a bar. We use labels L and R also for the massless fermions to keep the notation uniform. These fermions are related to the components of the Majorana-Weyl spinors of appendix~\ref{app:spinors-and-grassmanns} by
\begin{equation}
  \theta^{\sL\,j} = \theta_{1j} , \qquad
  \theta^{\sR\,j} = \bar{\theta}_{2j} .
\end{equation}
The fermions satisfy canonical anti-commutation relation of the form
\begin{equation}
  \big\{\bar{\theta}^{\sL\,j}(x),\theta^{\sL\,j}(y)\big\}=\delta(x-y) , \qquad
  \big\{\bar{\theta}^{\sR\,j}(x),\theta^{\sR\,j}(y)\big\}=\delta(x-y) ,
\end{equation}
for all masses~$|m_j|$.
For completeness we rewrite the supercharges from section~\ref{sec:supercurrents} in terms of $\theta^{\sL\,j}$ and $\theta^{\sR,j}$ as
\begin{equation}\label{eq:supercharges-quadr-fields}
  \begin{aligned}
    \QL = \frac{e^{-i\, \pi/4}}{2}\int d\sigma \smash{\Big(}
    & +2 P_{\bar{Z}} \theta^{\sL \, 4} + Z' (i\tilde{q} \bar{\theta}^{\sR \, 4} - q \theta^{\sL \, 4}) + i Z \theta^{\sL\, 4} \\
    &-2i P_{{Y}} \bar{\theta}^{\sL \, 3} - \bar{Y}' (\tilde{q} \theta^{\sR \, 3} - i q \bar{\theta}^{\sL \, 3}) - \alpha \bar{Y} \bar{\theta}^{\sL\, 3}  \\
    &-2i P_{{X}} \bar{\theta}^{\sL \, 2} - \bar{X}' (\tilde{q} \theta^{\sR \, 2} - i q \bar{\theta}^{\sL \, 2}) - (1-\alpha) \bar{X} \bar{\theta}^{\sL\, 2} \\
    &-2i P_{{W}} \bar{\theta}^{\sL \,1} - \bar{W}' (\tilde{q} \theta^{\sR\,1} - i q \bar{\theta}^{\sL \,1}) 
    \smash{\Big)},
    \\
    \QR = \frac{e^{-i\, \pi/4}}{2}\int d\sigma \smash{\Big(}
    & +2 P_{{Z}} \theta^{\sR \, 4} + \bar{Z}' (i\tilde{q} \bar{\theta}^{\sL \, 4} + q \theta^{\sR \, 4}) + i \bar{Z} \theta^{\sR\, 4} \\
    &-2i P_{\bar{Y}} \bar{\theta}^{\sR \, 3} - Y' (\tilde{q} \theta^{\sL \, 3} + i q \bar{\theta}^{\sR \, 3}) - \alpha Y \bar{\theta}^{\sR\, 3}  \\
    &-2i P_{\bar{X}} \bar{\theta}^{\sR \, 2} - X' (\tilde{q} \theta^{\sL \,2} + i q \bar{\theta}^{\sR \, 2}) - (1-\alpha) X \bar{\theta}^{\sR\, 2} \\
    &-2i P_{\bar{W}} \bar{\theta}^{\sR \,1} - W' (\tilde{q} \theta^{\sL\,1} + i q \bar{\theta}^{\sR \,1}) 
    \smash{\Big)}.
  \end{aligned}
\end{equation}
Note that $\QL$ is related to $\QR$ by exchanging a boson with its conjugate, swapping the labels L and R on the fermions and flipping the sign of the NS-NS flux coefficient, $q\to -q$. This is a manifestation of left-right symmetry.
From the equations above one can already expect the fields with mass $|m|=1-\alpha, \, \alpha, \, 0$ to be organised into representations with the same grading, and the representations with $|m|=1$ to have opposite grading.

\subsection{Expressions in terms of oscillators}
In order to introduce oscillators, we have defined the wave-function parameters~\eqref{eq:wavefun} which we repeat here for convenience:
\begin{equation}
  \begin{aligned}
    g_{\sL} (p,m_j) &= -\frac{\tilde{q} \, p}{2f_{\sL}(p,m_j)},
    \quad &
    g_{\sR} (p,m_j) &= -\frac{\tilde{q} \, p}{2f_{\sR}(p,m_j)},
    \\
    f_{\sL}(p,m_j) &= \sqrt{\frac{|m_j|+ q \, p +\omega_{\sL}(p,m_j)}{2}},
    \quad &
    f_{\sR}(p,m_j) &= \sqrt{\frac{|m_j|- q \, p +\omega_{\sR}(p,m_j)}{2}},
    \\
    \omega_{\sL}(p,m_j) &= \sqrt{p^2 + 2 \, |m_j|\, q\, p + m_j^2},
    \quad &
    \omega_{\sR}(p,m_j) &= \sqrt{p^2 - 2 \, |m_j|\, q\, p + m_j^2}.
  \end{aligned}
\end{equation}
These satisfy the useful identities
\begin{equation}
  \label{eq:wfunproperties}
  \begin{aligned}
    f_{\sL}(-p,m_j) &= +f_{\sR}(+p,m_j) , \qquad &
    f_{\sL}(p,m_j)^2 + g_{\sL}(p,m_j)^2 &= \omega_{\sL}(p,m_j),
    \\
    g_{\sL}(-p,m_ju) &= -g_{\sR}(+p,m_j),
    \qquad &
    f_{\sR}(p,m_j)^2 + g_{\sR}(p,m_j)^2 &= \omega_{\sR}(p,m_j).
  \end{aligned}
\end{equation}
We define the bosons as
\begin{equation}
  \begin{aligned}
    X_j &= \frac{1}{\sqrt{2\pi}} \int dp \left( \frac{1 }{\sqrt{\omega_{\sL}(p,m_j)}} \, a_{\sL\, j}^\dagger(p)\ e^{-i\, p\sigma} + \frac{1}{\sqrt{\omega_{\sR}(p,m_j)}}\, a_{\sR\, j}(p)  \ e^{i\, p\sigma} \right),
    \\
    \bar{X}_j &= \frac{1}{\sqrt{2\pi}} \int dp \left( \frac{1 }{\sqrt{\omega_{\sR}(p,m_j)}}\, a_{\sR\, j}^\dagger(p) \ e^{-i\, p\sigma}  + \frac{1}{\sqrt{\omega_{\sL}(p,m_j)}}\, a_{\sL\, j}(p)  \ e^{i\, p\sigma}\right),
    \\
    P_{\bar{X}_j} &= \frac{i}{\sqrt{2\pi}} \int \frac{dp}{2} \left( \sqrt{\omega_{\sL}(p,m_j)}\, a_{\sL\, j}^\dagger(p)  \ e^{-i\, p\sigma} - \sqrt{\omega_{\sR}(p,m_j)}\, a_{\sR\, j}(p)  \ e^{i\, p\sigma} \right),
    \\
    P_{X_j} &= \frac{i}{\sqrt{2\pi}} \int \frac{dp}{2} \left( \sqrt{\omega_{\sR}(p,m_j)}\, a_{\sR\, j}^\dagger(p)  \ e^{-i\, p\sigma}- \sqrt{\omega_{\sL}(p,m_j)}\, a_{\sL\, j}(p)  \ e^{i\, p\sigma}  \right),
  \end{aligned}
\end{equation}
where we introduced obvious short-hand notations $X_4=Z,$ $X_3=Y,$ $X_2=X$ and $X_1=W$.

We denote the fermionic annihilation operators by $d_{\sL\, j}, d_{\sR\, j}$ and  the creation operators are $d_{\sL\, j}^\dagger, d_{\sR\, j}^\dagger$ with $|m_j|=(0,1-\alpha,\alpha,1)$. Then
\begin{equation}
  \begin{aligned}
    \theta^{\sL \, j} &= \frac{e^{-i\, \pi/4}}{\sqrt{2\pi}} \int dp \left( \frac{g_{\sR}(p,m_j)}{\sqrt{\omega_{\sR}(p,m_j)}} \, d_{\sR\, j}^\dagger \ e^{-i\, p\sigma} -
      \frac{f_{\sL}(p,m_j)}{\sqrt{\omega_{\sL}(p,m_j)}} \, d_{\sL\, j} \ e^{i\, p\sigma} \right),
    \\
    \theta^{\sR \, j} &= \frac{e^{-i\, \pi/4}}{\sqrt{2\pi}} \int dp \left( \frac{g_{\sL}(p,m_j)}{\sqrt{\omega_{\sL}(p,m_j)}} \, d_{\sL\, j}^\dagger \ e^{-i\, p\sigma} -
      \frac{f_{\sR}(p,m_j)}{\sqrt{\omega_{\sR}(p,m_j)}} \, d_{\sR\, j} \ e^{i\, p\sigma} \right).
  \end{aligned}
\end{equation}
These definitions are such that the raising and lowering operators satisfy canonical (anti)commutation relations, which follows from the ones of the fields. Much like in references~\cite{Borsato:2014hja, Lloyd:2014bsa}  we can now use these definitions and the relations~\eqref{eq:wfunproperties} to rewrite the supercharges. It is tedious but straightforward to show that these take the form~\eqref{eq:supercharges}.

\section{Supercharges in the \texorpdfstring{$\alpha\to 1$}{alpha to 1} limit}
\label{app:a-to-1-limit}

Our construction bears some similarities to the one performed in references~\cite{Borsato:2014hja, Lloyd:2014bsa} for $\AdS_3\times\Sphere^3\times\Torus^4$, which is not surprising as $\AdS_3\times\Sphere^3\times\Sphere^3\times\Sphere^1$ takes that form---up to suitably compactifying the flat directions---when either sphere blows up. This can be achieved by sending $\alpha\to0$ or $\alpha\to1$. We have chosen our conventions such that, in the former case, the coordinates can be matched with the ones of references~\cite{Borsato:2014hja, Lloyd:2014bsa} as
\begin{equation}
  \begin{aligned}
    z_i &\atoone z_i,  && i=1,2 ,&
    \qquad
    y_i &\atoone y_i, \quad i=3,4 ,
    \\
    x_i &\atoone x_i, && i=6,7 ,&
    \qquad
    \psi &\atoone x_8, \quad 
    w \atoone x_9 .
  \end{aligned}
\end{equation}
Similarly, we can match the fermions as
\begin{equation}
  \begin{aligned}
    \theta^{\sL\, 4} &\atoone \eta_{\sL}^{\ 1} , \qquad
    &\theta^{\sL\, 3} &\atoone \eta_{\sL}^{\ 2} , \qquad
    &\theta^{\sL\, 2} &\atoone i \bar{\chi}_{+2} ,\qquad
    &\theta^{\sL\, 1} &\atoone  \bar{\chi}_{+1} ,
    \\
    \theta^{\sR\, 4} &\atoone \eta_{\sR \,1} ,\qquad
    &\theta^{\sR\, 3} &\atoone \eta_{\sR\, 2} ,\qquad
    &\theta^{\sR\, 2} &\atoone i \bar{\chi}_{-1} ,\qquad
    &\theta^{\sR\, 1} &\atoone  \bar{\chi}_{-2} ,
  \end{aligned}
\end{equation}
With these identifications, the supercharges match as
\begin{equation}
  \QL \atoone {\QL}^{1},
  \qquad
  \QR \atoone   \genQ_{\sR 1},
  \qquad
  \QbL \atoone \overline{\gen{Q}}_{\sL 1},
  \qquad
  \QbR \atoone {{\overline{\genQ}_{\sR}}}^{1} .
\end{equation}

\section{\texorpdfstring{$\algPSU(1|1)^2_{\ce}$}{psu(1|1) x psu(1|1)}-invariant S-matrices}
\label{app:Smat}

We collect the S-matrices invariant under $\algPSU(1|1)^2_{\ce}$ that are relevant for our results.
Although we do not write it explicitly, the dependence of the Zhukovski variables on a generic mass is always assumed. In particular, the results in this appendix are valid for any choice of the masses $|m|$ and $|m|'$ associated to the excitations with momenta $p$ and $q$, respectively.
To keep our notation simple, we denote a boson by $\phi$ and a fermion by $\psi$.

\paragraph{Same LR flavour}
If we decide to scatter two excitations both belonging to the representation $\varrho_{\sL}$ we find
\begin{equation}\label{eq:su(1|1)2-Smat-grad1}
\begin{aligned}
\mathcal{S}^{\sL\sL} \ket{\phi_p^{\sL} \phi_q^{\sL}} &= \phantom{+} A_{pq}^{\sL\sL} \ket{\phi_q^{\sL} \phi_p^{\sL}},
\qquad
&\mathcal{S}^{\sL\sL} \ket{\phi_p^{\sL} \psi_q^{\sL}} &= B_{pq}^{\sL\sL} \ket{\psi_q^{\sL} \phi_p^{\sL}} +  C_{pq}^{\sL\sL} \ket{\phi_q^{\sL} \psi_p^{\sL}}, \\
%%%
\mathcal{S}^{\sL\sL} \ket{\psi_p^{\sL} \psi_q^{\sL}} &= \phantom{+} F_{pq}^{\sL\sL} \ket{\psi_q^{\sL} \psi_p^{\sL}},\qquad
&\mathcal{S}^{\sL\sL} \ket{\psi_p^{\sL} \phi_q^{\sL}} &= D_{pq}^{\sL\sL} \ket{\phi_q^{\sL} \psi_p^{\sL}} +  E_{pq}^{\sL\sL} \ket{\psi_q^{\sL} \phi_p^{\sL}},
\end{aligned}
\end{equation}
The coefficients appearing are determined up to an overall factor. As a convention we decide to normalise $A_{pq}^{\sL\sL}=1$ and we find
\begin{equation}
\begin{aligned}
A^{\sL\sL}_{pq} &= 1, &
\qquad
B^{\sL\sL}_{pq} &= \phantom{-}\left( \frac{x^-_{\sL\, p}}{x^+_{\sL\, p}}\right)^{1/2} \frac{x^+_{\sL\, p}-x^+_{\sL\, q}}{x^-_{\sL\, p}-x^+_{\sL\, q}}, \\
C^{\sL\sL}_{pq} &= \left( \frac{x^-_{\sL\, p}}{x^+_{\sL\, p}} \frac{x^+_{\sL\, q}}{x^-_{\sL\, q}}\right)^{1/2} \frac{x^-_{\sL\, q}-x^+_{\sL\, q}}{x^-_{\sL\, p}-x^+_{\sL\, q}} \frac{\eta_{\sL\, p}}{\eta_{\sL\, q}}, 
\qquad &
D^{\sL\sL}_{pq} &= \phantom{-}\left(\frac{x^+_{\sL\, q}}{x^-_{\sL\, q}}\right)^{1/2}  \frac{x^-_{\sL\, p}-x^-_{\sL\, q}}{x^-_{\sL\, p}-x^+_{\sL\, q}}, \\
E^{\sL\sL}_{pq} &= \frac{x^-_{\sL\, p}-x^+_{\sL\, p}}{x^-_{\sL\, p}-x^+_{\sL\, q}} \frac{\eta_{\sL\, q}}{\eta_{\sL\, p}}, 
\qquad &
F^{\sL\sL}_{pq} &= - \left(\frac{x^-_{\sL\, p}}{x^+_{\sL\, p}} \frac{x^+_{\sL\, q}}{x^-_{\sL\, q}}\right)^{1/2} \frac{x^+_{\sL\, p}-x^-_{\sL\, q}}{x^-_{\sL\, p}-x^+_{\sL\, q}}.
\end{aligned}
\end{equation}
The S~matrix $\mathcal{S}^{\sR\sR}$ scattering two excitations that are both in the representation $\varrho_{\sR}$ is parameterised by scattering elements $A^{\sR\sR}_{pq}, B^{\sR\sR}_{pq},$ etc., obtained by substituting all labels left with labels right in the equations above.

If we scatter two excitations both transforming under $\widetilde{\varrho}_{\sL}$ we find an S~matrix that is related to the previous one. After choosing a convenient normalisation we write it as
\begin{equation}\label{eq:su(1|1)2-Smat-grad2}
\begin{aligned}
\mathcal{S}^{\tilde{\sL}\tilde{\sL}} \ket{\tilde{\phi}^{\sL}_p \tilde{\phi}^{\sL}_q} &= -F_{pq}^{\sL\sL} \ket{\tilde{\phi}^{\sL}_q \tilde{\phi}^{\sL}_p}, 
\qquad
&\mathcal{S}^{\tilde{\sL}\tilde{\sL}} \ket{\tilde{\phi}^{\sL}_p \tilde{\psi}^{\sL}_q} &= D_{pq}^{\sL\sL} \ket{\tilde{\psi}^{\sL}_q \tilde{\phi}^{\sL}_p}  -E_{pq}^{\sL\sL} \ket{\tilde{\phi}^{\sL}_q \tilde{\psi}^{\sL}_p}, \\
\mathcal{S}^{\tilde{\sL}\tilde{\sL}} \ket{\tilde{\psi}^{\sL}_p \tilde{\psi}^{\sL}_q} &= -A_{pq}^{\sL\sL} \ket{\tilde{\psi}^{\sL}_q \tilde{\psi}^{\sL}_p}, 
\qquad
&\mathcal{S}^{\tilde{\sL}\tilde{\sL}} \ket{\tilde{\psi}^{\sL}_p \tilde{\phi}^{\sL}_q} &= B_{pq}^{\sL\sL} \ket{\tilde{\phi}^{\sL}_q \tilde{\psi}^{\sL}_p}  -C_{pq}^{\sL\sL} \ket{\tilde{\psi}^{\sL}_q \tilde{\phi}^{\sL}_p}.
\end{aligned}
\end{equation}
The other cases to consider involve scattering of representations with different grading
\begin{equation}\label{eq:su(1|1)2-Smat-grad3}
\begin{aligned}
\mathcal{S}^{{\sL}\tilde{\sL}} \ket{\phi^{\sL}_p \tilde{\phi}^{\sL}_q} &= \phantom{+}B_{pq}^{\sL\sL} \ket{\tilde{\phi}^{\sL}_q \phi^{\sL}_p} -C_{pq}^{\sL\sL} \ket{\tilde{\psi}^{\sL}_q \psi^{\sL}_p},
\qquad
&\mathcal{S}^{{\sL}\tilde{\sL}} \ket{\phi^{\sL}_p \tilde{\psi}^{\sL}_q} &= \phantom{+}A_{pq}^{\sL\sL} \ket{\tilde{\psi}^{\sL}_q \phi^{\sL}_p} , \\
%%%
\mathcal{S}^{{\sL}\tilde{\sL}} \ket{\psi^{\sL}_p \tilde{\psi}^{\sL}_q} &= -D_{pq}^{\sL\sL} \ket{\tilde{\psi}^{\sL}_q \psi^{\sL}_p}+E_{pq}^{\sL\sL} \ket{\tilde{\phi}^{\sL}_q \phi^{\sL}_p} ,
\qquad
&\mathcal{S}^{{\sL}\tilde{\sL}} \ket{\psi^{\sL}_p \tilde{\phi}^{\sL}_q} &= -F_{pq}^{\sL\sL} \ket{\tilde{\phi}^{\sL}_q \psi^{\sL}_p} ,
\end{aligned}
\end{equation}
\begin{equation}\label{eq:su(1|1)2-Smat-grad4}
\begin{aligned}
\mathcal{S}^{\tilde{\sL}{\sL}} \ket{\tilde{\phi}^{\sL}_p \phi^{\sL}_q} &= \phantom{+}D_{pq}^{\sL\sL} \ket{\phi^{\sL}_q \tilde{\phi}^{\sL}_p} +E_{pq}^{\sL\sL} \ket{\psi^{\sL}_q \tilde{\psi}^{\sL}_p},
\qquad
&\mathcal{S}^{\tilde{\sL}{\sL}} \ket{\tilde{\phi}^{\sL}_p \psi^{\sL}_q} &= -F_{pq}^{\sL\sL} \ket{\psi^{\sL}_q \tilde{\phi}^{\sL}_p} , \\
%%%
\mathcal{S}^{\tilde{\sL}{\sL}} \ket{\tilde{\psi}^{\sL}_p \psi^{\sL}_q} &= -B_{pq}^{\sL\sL} \ket{\psi^{\sL}_q \tilde{\psi}^{\sL}_p}-C_{pq}^{\sL\sL} \ket{\phi^{\sL}_q \tilde{\phi}^{\sL}_p} ,
\qquad
&\mathcal{S}^{\tilde{\sL}{\sL}} \ket{\tilde{\psi}^{\sL}_p \phi^{\sL}_q} &= \phantom{+}A_{pq}^{\sL\sL} \ket{\phi^{\sL}_q \tilde{\psi}^{\sL}_p} .
\end{aligned}
\end{equation}
The matrices $\mathcal{S}^{\tilde{\sR}\tilde{\sR}},\mathcal{S}^{{\sR}\tilde{\sR}},\mathcal{S}^{\tilde{\sR}{\sR}}$ are found again by sending the labels $\sL \to \sR$ in the equations above, including in the spectral parameters.

\paragraph{Opposite LR flavour}
Scattering excitations carrying opposite LR flavour yields different results. To start, the scattering of $\varrho_{\sL}$ and $\varrho_{\sR}$ is
\begin{equation}\label{eq:su(1|1)2-Smat-LRgrad1}
\begin{aligned}
\mathcal{S}^{\sL\sR} \ket{\phi^{\sL}_p \phi^{\sR}_q} &= A^{\sL\sR}_{pq} \ket{\phi^{\sR}_q \phi^{\sL}_p} + B^{\sL\sR}_{pq} \ket{\psi^{\sR}_q \psi^{\sL}_p}, \qquad 
&\mathcal{S}^{\sL\sR} \ket{\phi^{\sL}_p \psi^{\sR}_q} &= C^{\sL\sR}_{pq} \ket{\psi^{\sR}_q \phi^{\sL}_p} , \\
\mathcal{S}^{\sL\sR} \ket{\psi^{\sL}_p \psi^{\sR}_q} &= E^{\sL\sR}_{pq} \ket{\psi^{\sR}_q \psi^{\sL}_p}+F^{\sL\sR}_{pq} \ket{\phi^{\sR}_q \phi^{\sL}_p} ,  \qquad 
& \mathcal{S}^{\sL\sR} \ket{\psi^{\sL}_p \phi^{\sR}_q} &= D^{\sL\sR}_{pq} \ket{\phi^{\sR}_q \psi^{\sL}_p} .
\end{aligned}
\end{equation}
Scattering them in the opposite order corresponds to considering the matrix $\mathcal{S}^{\sR\sL}$, that is found by swapping the labels L$\leftrightarrow$R.
The scattering elements may be written as
\begin{equation}
\begin{aligned}
 A^{\sL\sR}_{pq} &= \zeta^{\sL\sR}_{pq}\left(\frac{x^+_{\sL\, p}}{x^-_{\sL\, p}} \right)^{1/2} \frac{1-\frac{1}{x^+_{\sL\, p} x^-_{\sR\,q}}}{1-\frac{1}{x^-_{\sL\, p} x^-_{\sR\,q}}}, 
 \qquad &
 B^{\sL\sR}_{pq} &= -\frac{2i}{h} \, \left(\frac{x^-_{\sL\, p}}{x^+_{\sL\, p}}\frac{x^+_{\sR\, q}}{x^-_{\sR\, q}} \right)^{1/2} \frac{\eta_{\sL\, p}\eta_{\sR\, q}}{ x^-_{\sL\, p} x^+_{\sR\, q}} \frac{\zeta^{\sL\sR}_{pq}}{1-\frac{1}{x^-_{\sL\, p} x^-_{\sR\, q}}}, \\
  C^{\sL\sR}_{pq} &= \phantom{-}\zeta^{\sL\sR}_{pq},
\qquad &
D^{\sL\sR}_{pq} &=\phantom{-}\zeta^{\sL\sR}_{pq}\left(\frac{x^+_{\sL\, p}}{x^-_{\sL\, p}}\frac{x^+_{\sR\, q}}{x^-_{\sR\, q}} \right)^{1/2} \frac{1-\frac{1}{x^+_{\sL\, p} x^+_{\sR\, q}}}{1-\frac{1}{x^-_{\sL\, p} x^-_{\sR\, q}}}, \\
 E^{\sL\sR}_{pq} &= - \zeta^{\sL\sR}_{pq}\left(\frac{x^+_{\sR\, q}}{x^-_{\sR\, q}} \right)^{1/2} \frac{1-\frac{1}{x^-_{\sL\, p} x^+_{\sR\, q}}}{1-\frac{1}{x^-_{\sL\, p} x^-_{\sR\, q}}},
\qquad &
F^{\sL\sR}_{pq} &= \frac{2i}{h} \, \left(\frac{x^+_{\sL\, p}}{x^-_{\sL\, p}}\frac{x^+_{\sR\, q}}{x^-_{\sR\, q}} \right)^{1/2}  \frac{\eta_{\sL\, p}\eta_{\sR\, q}}{ x^+_{\sL\, p} x^+_{\sR\, q}} \frac{\zeta^{\sL\sR}_{pq}}{1-\frac{1}{x^-_{\sL\, p} x^-_{\sR\, q}}},
\end{aligned}
\end{equation}
where we have multiplied the matrix by a convenient overall factor
\begin{equation}
\zeta^{\sL\sR}_{pq}=\left( \frac{x^+_{\sL\, p}}{x^-_{\sL\, p}} \right)^{-1/4}\left( \frac{x^+_{\sR\, q}}{x^-_{\sR\, q}} \right)^{-1/4}
\left( \frac{1-\frac{1}{x^-_{\sL\, p}x^-_{\sR\, q}}}{1-\frac{1}{x^+_{\sL\, p}x^+_{\sR\, q}}} \right)^{1/2},
\end{equation}
in such a way that unitarity is simply $\mathcal{S}^{\sL\sR}\mathcal{S}^{\sR\sL}=\mathbf{1}$.

As previously, we write also the other S~matrices corresponding to the other choices of the gradings of the representations
\begin{equation}\label{eq:su(1|1)2-Smat-LRgrad2}
\begin{aligned}
\mathcal{S}^{\tilde{\sL}\tilde{\sR}} \ket{\tilde{\phi}^{\sL}_p \tilde{\phi}^{\sR}_q} &= -E^{\sL\sR}_{pq} \ket{\tilde{\phi}^{\sR}_q \tilde{\phi}^{\sL}_p} + F^{\sL\sR}_{pq} \ket{\tilde{\psi}^{\sR}_q \tilde{\psi}^{\sL}_p}, \qquad 
&\mathcal{S}^{\tilde{\sL}\tilde{\sR}} \ket{\tilde{\phi}^{\sL}_p \tilde{\psi}^{\sR}_q} &= D^{\sL\sR}_{pq} \ket{\tilde{\psi}^{\sR}_q \tilde{\phi}^{\sL}_p} , \\
\mathcal{S}^{\tilde{\sL}\tilde{\sR}} \ket{\tilde{\psi}^{\sL}_p \tilde{\psi}^{\sR}_q} &= -A^{\sL\sR}_{pq} \ket{\tilde{\psi}^{\sR}_q \tilde{\psi}^{\sL}_p}+B^{\sL\sR}_{pq} \ket{\tilde{\phi}^{\sR}_q \tilde{\phi}^{\sL}_p} ,  \qquad 
& \mathcal{S}^{\tilde{\sL}\tilde{\sR}} \ket{\tilde{\psi}^{\sL}_p \tilde{\phi}^{\sR}_q} &= C^{\sL\sR}_{pq} \ket{\tilde{\phi}^{\sR}_q \tilde{\psi}^{\sL}_p} .
\end{aligned}
\end{equation}
\begin{equation}\label{eq:su(1|1)2-Smat-LRgrad3}
\begin{aligned}
\mathcal{S}^{\tilde{\sL}\sR} \ket{\tilde{\phi}^{\sL}_p \phi^{\sR}_q} &= +D^{\sL\sR}_{pq} \ket{\phi^{\sR}_q \tilde{\phi}^{\sL}_p},  \qquad 
& \mathcal{S}^{\tilde{\sL}\sR} \ket{\tilde{\phi}^{\sL}_p \psi^{\sR}_q} &= -E^{\sL\sR}_{pq} \ket{\psi^{\sR}_q \tilde{\phi}^{\sL}_p} -F^{\sL\sR}_{pq} \ket{\phi^{\sR}_q \tilde{\psi}^{\sL}_p}, \\
\mathcal{S}^{\tilde{\sL}\sR} \ket{\tilde{\psi}^{\sL}_p \psi^{\sR}_q} &= -C^{\sL\sR}_{pq} \ket{\psi^{\sR}_q \tilde{\psi}^{\sL}_p},  \qquad 
& \mathcal{S}^{\tilde{\sL}\sR} \ket{\tilde{\psi}^{\sL}_p \phi^{\sR}_q} &= +A^{\sL\sR}_{pq} \ket{\phi^{\sR}_q \tilde{\psi}^{\sL}_p} +B^{\sL\sR}_{pq} \ket{\psi^{\sR}_q \tilde{\phi}^{\sL}_p}.
\end{aligned}
\end{equation}
\begin{equation}\label{eq:su(1|1)2-Smat-LRgrad4}
\begin{aligned}
\mathcal{S}^{\sL\tilde{\sR}} \ket{\phi^{\sL}_p \tilde{\phi}^{\sR}_q} &= +C^{\sL\sR}_{pq} \ket{\tilde{\phi}^{\sR}_q \phi^{\sL}_p},  \qquad 
& \mathcal{S}^{\sL\tilde{\sR}} \ket{\phi^{\sL}_p \tilde{\psi}^{\sR}_q} &= +A^{\sL\sR}_{pq} \ket{\tilde{\psi}^{\sR}_q \phi^{\sL}_p} - B^{\sL\sR}_{pq} \ket{\tilde{\phi}^{\sR}_q \psi^{\sL}_p}, \\
\mathcal{S}^{\sL\tilde{\sR}} \ket{\psi^{\sL}_p \tilde{\psi}^{\sR}_q} &= -D^{\sL\sR}_{pq} \ket{\tilde{\psi}^{\sR}_q \psi^{\sL}_p},  \qquad 
& \mathcal{S}^{\sR\tilde{\sR}} \ket{\psi^{\sL}_p \tilde{\phi}^{\sR}_q} &= -E^{\sL\sR}_{pq} \ket{\tilde{\phi}^{\sR}_q \psi^{\sL}_p} +  F^{\sL\sR}_{pq} \ket{\tilde{\psi}^{\sR}_q \phi^{\sL}_p}.
\end{aligned}
\end{equation}

\bibliographystyle{ads3-s1-s-matrix-arxiv-v1}
\bibliography{ads3-s1-s-matrix-arxiv-v1}

\end{document}